\pdfoutput=1
\documentclass{article}
\UseRawInputEncoding
\usepackage{ijcai24}

\usepackage{times}
\usepackage{soul}

\usepackage{url}
\usepackage[hidelinks]{hyperref}
\usepackage[utf8]{inputenc}
\usepackage[small]{caption}
\usepackage{graphicx}
\usepackage{amsmath}
\usepackage{amsthm}
\usepackage{booktabs}
\usepackage[switch]{lineno}
\usepackage{amssymb}
\usepackage{subfigure}
\usepackage{stfloats}
\usepackage{cases}
\usepackage{multirow}
\usepackage{tabularx}
\usepackage{booktabs}
\usepackage[capitalize]{cleveref}
\usepackage{mathrsfs} 

\usepackage{algpseudocode}

\usepackage[linesnumbered,ruled,vlined]{algorithm2e}

\newtheorem{theorem}{Theorem}

\usepackage{amsthm} 

\newtheorem{lemma}{Lemma}
\newtheorem{definition}{Definition}

\title{LEAP: Optimization Hierarchical Federated Learning on Non-IID Data with Coalition Formation Game}

\author{
Jianfeng Lu$^{1,2}$\and
Yue Chen$^1$\and
Shuqin Cao$^{1,2}$ \footnote{Corresponding Author.}\and
Longbiao Chen$^3$\and
Wei Wang$^{1,2}$\And
Yun Xin$^1$\\
\affiliations
$^1$School of Computer Science and Technology, Wuhan University of Science and Technology, China\\
$^2$Hubei Province Key Laboratory of Intelligent Information Processing and Real-time Industrial System, Wuhan University of Science and Technology, China\\
$^3$School of Informatics, Xiamen University, China\\
\emails
\{lujianfeng, chenyue, shuqincao\}@wust.edu.cn,
longbiaochen@xmu.edu.cn,
wangwei8@wust.edu.cn
yunxin.wust@gmail.com
}

\begin{document}

\maketitle

\begin{abstract}
Although Hierarchical Federated Learning (HFL) utilizes edge servers (ESs) to alleviate communication burdens, its model performance will be degraded by non-IID data and limited communication resources. Current works often assume that data is uniformly distributed, which however contradicts the heterogeneity of IoT. Solutions of additional model training to check the data distribution inevitably increases computational costs and the risk of privacy leakage. The challenges in solving these issues are how to reduce the impact of non-IID data without involving raw data and how to rationalize the communication resource allocation for addressing straggler problem. To tackle these challenges, we propose a novel optimization method based on coa\underline{L}ition formation gam\underline{E} and gr\underline{A}dient \underline{P}rojection, called LEAP. Specifically, we combine edge data distribution with coalition formation game innovatively to adjust the correlations between clients and ESs dynamically, which ensures optimal correlations. We further capture the client heterogeneity to achieve the rational bandwidth allocation from coalition perception and determine the optimal transmission power within specified delay constraints at client level. Experimental results on four real datasets show that LEAP is able to achieve 20.62\% improvement in model accuracy compared to the state-of-the-art baselines. Moreover, LEAP effectively reduce transmission energy consumption by at least about 2.24 times.
\end{abstract}

\section{Introduction}
As a novel distributed machine learning paradigm, FL \cite{FL_FedAvg} has gained the attention of many fields, such as the Internet of Things (IoT) \cite{IoT1}, smart transportation \cite{Smart1}, and healthcare \cite{healthcare1}, to break down the information silos while enabling privacy preservation. With the power of FL, Artificial Intelligence (AI) can effectively handle machine learning tasks involving decentralized data, which draws upon the advantages of distribution machine learning, while also significantly eliminates the privacy risks. During the training process, only model parameters are transmitted without involving local data, breaking information silos and greatly improving training efficiency \cite{ref2}. However, FL performance is affected by various factors, both in the training and transmission phases \cite{ref3}. During the training phase, FL involves hundreds and thousands of clients, and the data distribution of each client is severely different due to diverse user behavior patterns and data collection methods. Consequently, the local data of an individual client fails to represent the overall data distribution of the environment, leading to significant reductions in model performance, compromising the model's generalization capability. In the transmission phase, there is high communication latency and instability between clients and central server (CS) by mass and frequent data transfers, reducing model training efficiency and even increasing the risk of data leakage.

In recent years, numerous solutions have been proposed to address data heterogeneity and communication bottlenecks. For a FL task, client's data distribution is very important to the performance of FL model. When the data distribution of each client deviates seriously, the FL model is difficult to learn and its performance is low. For example, non-IID data causes a number of FL model performance issues, including decreased accuracy, sluggish model convergence, and model communication delays. Reinforcement learning \cite{RL} and data augmentation \cite{data_ugmentation} were proposed to address the non-IID challenges. These studies overemphasize the importance of individual clients, ignoring the performance improvement of the model benefited from the combination of local updates. In addition, despite their significance, most of these approaches, require auxiliary models or extra data transmissions in FL, potentially introducing additional complexities. To alleviate communication pressures, various techniques such as model compression \cite{model_compression}, gradient sparsity \cite{gradient_sparsification}, and over-the-air computation \cite{over-the-air} were proposed. Although these methods can effectively reduce communication overhead, they may still result in bottlenecks in communication at the CS each communication round. This is due to the fact that during some training epochs, the CS still receives large model weight updates.

Inspired by \cite{Client_Edge_Cloud}, we try to reduce the impact of non-IID distribution on HFL training by increasing the degree of IID of data distribution between ESs. Additionally, we would like to further optimize the resource allocation scheme, thus reducing the communication latency. The combination of these two objectives gives rise to an extremely complex and difficult problem, which leads to the following three challenges: First, \textit{different edge association relationships represent different edge data distribution.} Once edge association changes, the data distribution will evolve in an unpredictable direction, potentially reducing or increasing the degree of edge IID. Therefore, the impact of changes in association relationships on changes in data distribution is vague and uncertain. Second, \textit{straggler problem caused by worst-performing client.} Edge aggregation latency is susceptible to the communication performance of the worst-performing client in synchronized FL. Dynamic edge association relationships make it difficult to capture communication performance information for each edge coalition. As a result, targeted resource optimization is impossible. Third, \textit{contradiction between task execution latency requirements and clients' energy consumption.} Sufficient resource investment can meet the task requirements but may cause excessive overhead on clients, which is impractical. Consequently, striking a balance that satisfies the needs of both parties simultaneously poses a formidable challenge.
 
To tackle the abovementioned challenges, we propose an optimization method for HFL based on coa\underline{L}ition formation gam\underline{E} and gr\underline{A}dient \underline{P}rojection method, named LEAP, which not only effectively reduces the impact of cross-edge non-IID, but also improves the communication efficiency. The main contributions of our work are as follows:
\begin{itemize}
    \item Theoretically, we focus on the effect of multidimensional properties (i.e., time delay, energy consumption, and data distribution) on the performance of HFL. In LEAP, we transform data distribution optimization problem into edge correlation problem and further optimize heterogeneous resource problem.

    \item Methodologically, we construct a coalition formation game by analyzing the relationship between edge association and edge data distribution similarity. Moreover, we prove the existence of stable coalitions. Based on this, we utilize the gradient projection method to calculate the optimal bandwidth allocation for each coalition, and futher determine the transmission performance of heterogeneous clients to ensure that the latency requirements of the tasks are met.

    \item Experimentally, we validate the effectiveness of LEAP on four real datasets and baselines, it is able to achieve 20.62\% improvement in accuracy compare to mesn-shift algorithm. Moreover, experiments demonstrate that our optimization method is able to reduce the transmission energy consumption by at least 2.4 times while ensuring that the maximum latency requirement is met.
\end{itemize}

\section{Related Work}
In this section, we briefly discuss related work on non-IID data and communication bottleneck in FL.

\textbf{Non-IID Data.} In FL, the non-IID data, caused by the heterogeneity among clients, poses a challenge for training robust FL models due to its impact on slowing down the convergence of the global model \cite{ref1}. Many efforts have been made to address this issue. For instance, Arisdakessian \textit{et al.} \cite{coalition_federated_learning} proposed a trust-based coalitional FL approach, which mitigates non-IID problems by sharing data of coalition masters. Lu \textit{et al.} \cite{meanshift} utilizes the mean-shift algorithm to cluster client according to data distribution and then selects clients from different clusters to participate in training. Shin \textit{et al.} \cite{XORMixup} proposed a novel approach that uses XorMixup hybrid data enhancement technology to generate synthetic yet realistic sample data on the server to solve the issue of unbalanced training datasets in one-shot FL, but this approach will bring a large computational burden.

Data-sharing operation raises privacy concerns for clients, thus limiting its application scenarios and it is difficult to operate under privacy-preserving FL. The method of selecting clients through clustering does mitigate the non-IID problem, but it does not guarantee that the final selection result is optimal. In contrast, our work can find optimal edge association relationship in no additional model training without considering the raw data leakage problem.

\textbf{Communication Bottleneck.} In FL, communication cost is a significant factor that affects overall efficiency and effectiveness, while the uplink transmission rate of the underlying client is a major bottleneck in the training process. Many researchers proposed related solutions. For example, Mills \textit{et al.} \cite{Mills} focused on enhancing communication efficiency in FL by combining a distributed Adam optimizer with a compression technique. They emphasized reducing the uploaded data size during training rounds to mitigate communication costs. Building upon of model compression, Liu \textit{et al.} \cite{Liu} applied it to wireless FL to alleviate local computation and communication bottlenecks.

The abovementioned studies were conducted on cloud-based FL systems, whereby the CS receives the local model from the clients. However, in cloud-based FL, the transmission distance can often be considerable, resulting in unstable and undependable communication among the clients and CS. In our study, we make full use of abundant bandwidth resources of ESs and design communication optimization method for heterogeneous clients to solve the resource allocation problem in HFL, as well as to improve the effectiveness of HFL in heterogeneous client environments.

\section{System Model and Problem Formulation} In this section, we introduce the workflow of HFL, refine its multidimensional properties, and give an explicit definition of the optimization problem.

\subsection{HFL Framework}
We consider a HFL framework that consists of a set $ {\cal N} = \left\{ {1, \cdots, N} \right\} $ of clients , a set ${\cal M} = \left\{ {1, \cdots, M} \right\}$ of ESs, and a CS. The data set of the client $n$ is denoted as $\mathcal{D}_n =\left \{ \mathcal{X}_n ,\mathcal{Y}_n  \right \} $, where $\mathcal{X}_n =\left \{x_{n,1}, \cdots , x_{n,D_n} \right \} $ are the training dataset, $\mathcal{Y}_n =\left \{y_{n,1}, \cdots , y_{n,D_n} \right \}$ are the corresponding label set, and $|\mathcal{D}_n|$ is the number of training data owned by the client $n$. CS aims to train a model, with parameters denoted by a vector $\omega$, over $K$ iterations to minimize the global loss $ {L^K}\left( \omega  \right) $. The ESs, base stations in cellular networks or RSUs (Road Side Units) in vehicular networks, are employed to facilitate the uplink transmission of parameter updates by distributing orthogonal resource blocks to their clients. Then, each client can only associate with one ES to perform the model training. We define ${\cal G}_m$ as the set of clients that associate with ES $m$, i.e., ${\cal G}_m = \left \{ n\in {\cal N} : a_{m,n}=1\right \} $, where $\mathbf{A} = \left [ a \right ]_{M\times N}$ is the edge association matrix and ${{\cal G}_m} \cap {{\cal G}_{m'}} = \emptyset$ for $m \ne m'$. The HFL iteration $i$ consists of four main steps as follows \cite{HFL_step}:

\begin{itemize}
    \item \textit{Local Training:} Each client receives the intermediate model from ES $m$ denoted by $ {\omega}^{ i }$, to train a local model using its dataset.
    
    \item  \textit{Local Model Parameter Transmission:} After every $\tau_c$ rounds of local updates, clients transmits the updated local model ${\omega}_{n,m}^{i,\tau_c}$ to the associated ES $m$.
    
    \item \textit{Edge Aggregation:} ES $m$ aggregates the local model parameters from its associated clients to derive the intermediate model ${\omega}_m^{ {i,1} }$, which is transmitted back to the clients for the next edge iteration.
    
    \item \textit{Global Aggregation:} At the end of predefined intervals $\tau_e$, each ES transmits the intermediate model ${\omega} _m^{ {i , \tau_e } }$ to CS for aggregation to derive the updated global model ${\omega ^{ {i + 1 } }}$ and transmits new global model back to clients for the next global iteration.
\end{itemize}
The entire process described above will continue until a predetermined number of global training rounds $\tau_g$ is reached.

\subsection{Multi-Dimensional Properties in HFL}
The efficiency and sustainability of HFL is affected by execution time, energy cost and data quality. These three comprehensive properties consider the impacts of different aspects on FL systems. We hence give a formal definition of each property as follows:
\begin{definition}
The muti-dimensional properties of FL $\mathcal{V}$ is represented as a 3-tuple $\left( \cal{T},\cal{E}, \cal{J} \right)$, i.e., execution time $\cal T$, energy consumption $\cal E$, and data distribution similarity $\cal J$.

\begin{itemize}
	\item $\cal T$ means the execution time of a task, which includess computation latency $\cal{T}^C$ and communication latency $\cal{T}^U$, i.e., 
\begin{numcases}{}
	 {\cal T}_{n,t}^{\cal C} = \tau_c\frac{c_n|\mathcal{D}_n|}{f_n}, \label{时间_计算} \\
	 {\cal T}_{n,t}^{\cal U} =\frac{\mathbb{Z}}{\mathbb{R}_{n,m}} ,\label{时间_通信}\\
	 {\cal T}_{n,t} = {\cal T}_{n,t}^{\cal C} + {\cal T}_{n,t}^{\cal U}, \\
	 {\cal T}_m =\sum_{t=1}^{\tau_e } \left ( \underset{n\in \mathcal{G}_m }{\max} {\cal T}_{n,t}  \right ), \label{T_m}\\
	 {\cal T} =\tau_g \underset{ m\in \mathcal{M} }{\max}{\cal T}_m,
\end{numcases}
where $t$ means a edge iteration, ${c_n}$ is the number of CPU cycles for training unit data, $f_n$ is the CPU cycle frequency that determines the computational power, and $\mathbb{Z}$ is the model size. Clients upload local models to ESs via frequency domain multiple access (FDMA). The bandwidth allocation matrix is defined as $\mathbf{B} = \left [ B \right ]_{1\times M}$. $B_m$ is the bandwidth allocated for ES $m$ and $B_{m,n}^U$ is the bandwidth allocated for client $n$ to upload local model. $\mathbb{R}_{n,m}$ represents the uplink transmission rate of client $n$:
\begin{equation}
\mathbb{R}_{n,m}=B_{n,m}^U\log_{2}\left ( {1+\frac{p_{n,m}h_{n,m} }{B_{n,m}^U \mathbb{N}_0}}   \right ),
\label{上行链路}
\end{equation}
where $p_{n,m}$ denotes the transmission power of the client $n$, $h_{n,m}$ is the channel gain between client $n$ and ES $m$, and $\mathbb{N}_0$ is the power of additive white Gaussian noise.  
	\item $\cal E$ defines the energy consumption, which contains the training energy ${\cal E}^{\cal C}$ and transmission energy ${\cal E}^{\cal U}$.

\begin{numcases}{}
	 {\cal E}_{n,t}^{\cal C} = \tau_c \varphi {c_n}|\mathcal{D}_n|f_n^2, \label{能耗_计算}\\
	 {\cal E}_{n,t}^{\cal U} = {\cal T}_{n,t}^{\cal U}{p_{n}}, \label{能耗_通信}\\
	 {\cal E}_{n,t} = {\cal E}_{n,t}^{\cal C} + {\cal E}_{n,t}^{\cal U}, \label{能耗_t} \\
	 {\cal E}_m = \sum_{n\in \mathcal{G} _m} \tau_g \tau_e {\cal E}_{n,t}, \\
	 {\cal E}=\sum_{m=1}^{M} {\cal E}_m, \label{能耗_总}
\end{numcases}
where $\varphi$ is the effective capacitance parameter of the computing chipset.

	\item $\cal J$ denotes the data distribution similarity cross-edge that measured by Jensen Shannon Divergence (JSD). JSD and data distribution similarity are negatively correlated. A lower JSD implies that the two sets of data are more likely to fulfill the assumption of being IID. We only compute the JSD value once between the two ESs because of the symmetry of JSD, i.e.,  $\mathcal{JS}(Q_a,Q_b)=\mathcal{JS}(Q_b,Q_a)$ and $JSD\in \left [ 0,1 \right ] $. 
	\begin{numcases}{}
	 \mathcal{JS}(Q_a,Q_b)=\frac{\mathcal{KL}(Q_a,\mathbb{M}_{a,b})+\mathcal{KL}(Q_b,\mathbb{M}_{a,b})}{2}, \label{JS散度计算}\\
	 \mathcal{\overline{JS}} = \frac{\sum_{i=1}^{M-1} \sum_{j=i+1}^{M} \mathcal{JS}(Q_i, Q_j)}{M}, \label{JS散度计算_Avg}
	\end{numcases}
where $Q_a$ and $Q_b$ are the probability distributions of the data under ES $a$ and ES $b$, $\mathcal{KL}(\cdot)$ denotes the KLD (Kullback-Leibler Divergence) \cite{KLD}. $M_{a,b}$ denotes the mean distribution of $Q_a$ and $Q_b$:
\begin{equation}
    \mathbb{M}_{a,b}=\frac{Q_a+Q_b}{2}.
    \label{JS散度计算_M}
\end{equation}
\end{itemize}
\end{definition}

\paragraph{Remark:} We mainly consider the case of equal number of local data in this paper. And we are based on synchronized FL scenario, so the execution latency in a round of global iteration under a ES depends on the last completed client, as shown in  Eq. (\ref{T_m}). Due to the high transmission power of ES, the aggregation time and downlink transmission time are ignored compared to the local training and upload time. The clients under the same ES is share bandwidth resources equally, which means that $B_{1,m}^U=\cdots =B_{n,m}^U,n\in \mathcal{G} _m$.

\begin{figure}[t]
 \includegraphics[scale=0.3]{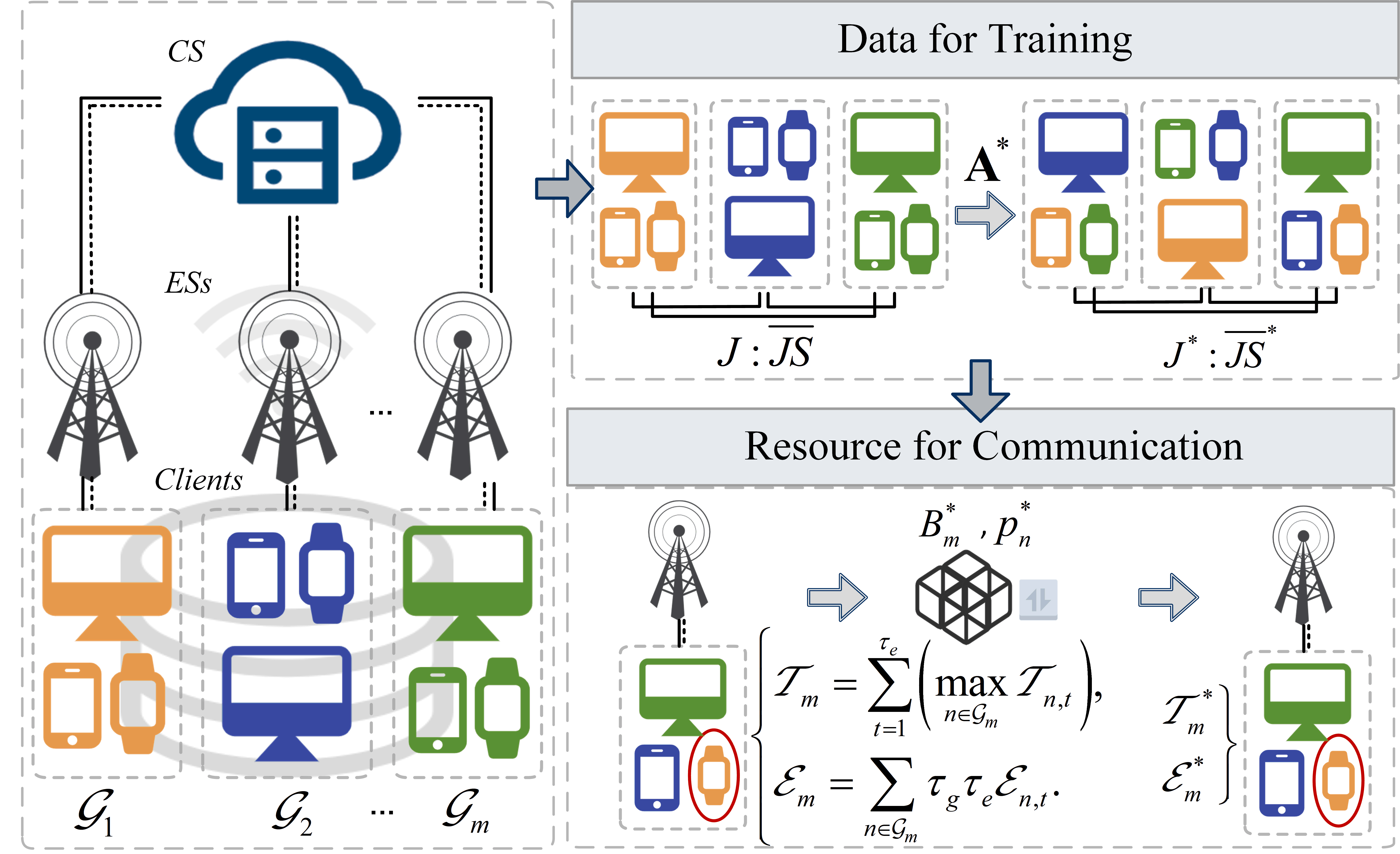}
\caption{An overview of LEAP}
\label{system_model}
 \vspace{-0.6em}
\end{figure}

\subsection{Problem Formulation}
For a FL task,  high-quality data can enhance model performance, while longer time delays and higher energy consumption bring negative impacts. For ease of representation, we define the utility of the network as a function of $\cal J$ and $\cal E$.
\begin{equation}
    {\cal U} = \lambda_1 (1- \mathcal{\overline{JS}}) - \lambda _2 {\cal E},
\end{equation}
where $ \lambda_1$ and $ \lambda_2$ are weighting parameters. Based on the multi-dimensional properties, we model the problem as:
\begin{subequations}
\begin{align}
{\mathbb{P}_1 }:& \quad  \underset{\bf{A,B,p}}{\max}\enspace {\cal U} \quad \& \quad \underset{\bf{A,B,p}}{\min} \enspace {\cal T},  \label{eq:Max&Min}\\
& \quad \text{s.t.}\quad a_{n,m}\in \left \{ 0,1 \right \}, \forall n\in \mathcal{N}, \forall m\in \mathcal{M}, \label{eq:constraint1} \\ 
& \quad \quad \quad  \sum_{m=1}^{M} B_m=B, \forall m\in \mathcal{M}, \label{eq:constraint2} \\
& \quad \quad \quad  B_m > 0, \forall m\in \mathcal{M}, \label{eq:constraint3} \\
& \quad \quad \quad  p_n\in \left ( 0,p_n^{max} \right ], \forall n\in \mathcal{N}, \label{eq:constraint4} \\
& \quad \quad \quad  {\cal T}_{n,t}\le \frac{\mathbb{I}}{\tau_e\tau_g}, \forall n\in \mathcal{N}, \label{eq:constraint5}
\end{align}
\end{subequations}
where Eq. (\ref{eq:constraint1}) indicates that each client can only associate with one ES at a time. Eq. (\ref{eq:constraint2}) and Eq. (\ref{eq:constraint3}) are the bandwidth constraints of uplink channels. The transmission power constraint is given by Eq. (\ref{eq:constraint4}) and Eq. (\ref{eq:constraint5}). $\mathbb{I}$ in Eq. (\ref{eq:constraint5}) is the maximum execution latency of the currently executed FL task.

LEAP decomposes the original problem $\mathbb{P}_1$ into several subproblems, as shown in Fig.\ref{system_model}, which can be solved one by one by combining coalition formation game and gradient projection method. In the coalition formation game, we design a coalition-friendly preference rule $\cal F$ to determine the optimal edge association relationship $\textbf{A}^*$. Based on stable coalition structure, in order to reduce the communication delay and energy consumption, LEAP optimizes the bandwidth allocation $\textbf{B}^*$ using the gradient projection method. In addition to this, LEAP captures the heterogeneous client communication resource conditions and determines the optimal transmission power $\textbf{p}^*$ of the clients based on satisfying the maximum delay of the task.

\section{Optimal Solution Based on LEAP}
In this section, we address the previously formulated problem $\mathbb{P}_1$. We start by identifying a stable coalition partition, and then optimize bandwidth allocation and transmission power based on this result.

\subsection{Optimal Data Distribution}
To achieve the goal of optimal network utility, or in other words, to obtain a lower result of JSD \cite{JSD}, we need to enhance the similarity of data distribution between ESs. Coalition game possesses an excellent tool for revealing the coalition formulation process. We model the problem of minimizing the JSD value of data distribution among ESs as a coalition formation game.

\begin{definition}
    A coalition formation game $\cal C$ is represented as a 4-tuple ($\cal N$, $\cal O$, $\cal F$, $\bf{A}$), i.e., a set $\cal N$ of clients, a coalition partition $\cal O$, a preference relation $\cal F$, and a game strategy profile $\bf{A}$.
    \begin{itemize}
        \item $\cal N$: A set ${\cal N} = \left \{ 1, \cdots, N  \right \} $ of players.

	   \item $\cal O$: A coalition partition ${\cal O} = \left\{ {{\cal G}} \right\}_{1}^{M}$, where coalition ${{\cal G}_m} \subseteq {\cal O}$, $ \cup _{m = 1}^{\rm{M}}{{\cal G}_m} = {\cal N}$ and $m$ denotes the index of the coalition or ES.

        \item $\cal F$: A preference relation  ${\succ  _n}$ is a complete, reflexive, and transitive binary relation over the set of all coalitions that client $n$ may join in, i.e., ${{\cal G}_1}{ \succ _n}{{\cal G}_2}$ indicates that client $n$ strictly prefers joining coalition ${{\cal G}_1}$ over coalition ${{\cal G}_2}$.
       
        \item $\bf{A}$: A strategy profile of edge association of clients, i.e., $a_{m,n}=1$ means client $n$ associates with ES $m \in {\cal M}$.
    \end{itemize}
\end{definition}
When determining client preferences for multiple coalitions, the order of coalition preferences can be determined according to different rules. For example, the ``Selfishness Rule'', which only considers individual's choices, and the ``Pareto Rule'', which never harms the choices of any member in original coalition and new coalition, but both are too extreme \cite{preference_rule}. The former completely ignores the development of other clients in same coalition, posing a risk of harming coalition partiton. The latter is too strict for the development of clients and coalitions. To minimize $\mathcal{\overline{JS}}$, we use a preference rule that places greater emphasis on the collective welfare of the entire coalition, called coalition-friendly preference rule, which the definition is as follows,
\begin{definition}
If there are two potential coalitions that client $n$ can join, i.e., ${{\cal G}_a, {\cal G}_b} \in {\cal O}$, then the preference relation is
\begin{equation}
\mathcal{G}_a\succ _n \mathcal{G}_b\Leftrightarrow  \mathcal{\overline{JS}}_{\mathcal{G}_b \to\mathcal{G}_a}^n<  \mathcal{\overline{JS}}_{\mathcal{G}_b}^n,
\label{客户偏好关系}
\end{equation}
where $ \mathcal{\overline{JS}}_{\mathcal{G}_b \to\mathcal{G}_a}^n$ means the $ \mathcal{\overline{JS}}$ value after client $n$ leaves original coalition $\mathcal{G}_b$ to join the new coalition $\mathcal{G}_a$, and $ \mathcal{\overline{JS}}_{\mathcal{G}_b}^n$ means the $ \mathcal{\overline{JS}}$ value before client $n$ leaves.
\end{definition}

The coalition-friendly preference rule fits well with our requirement, i.e., client towards a globally optimal solution by considering the reduction of $ \mathcal{\overline{JS}}$ before and after the switch. Based on the preference relations given in Eq. (\ref{客户偏好关系}), we define the switch rule in the coalition formation game:
\begin{definition}\label{定义_Switch rule}
    Given a partition ${\cal O} = \left\{ {{\cal G}} \right\}_{1}^{M}$, the client $n \in {{\cal G}_a}$ decides to leave the original coalition ${{\cal G}_a}$ and move to another coalition ${{\cal G}_{\rm{b}}}$, $b \ne a$ , if and only if ${{\cal G}_{\rm{b}}} \cup \left\{ n \right\}{\underline  \succ  _n}{{\cal G}_{\rm{a}}}$. The new coalition partition can be described as $\widetilde {\cal O} \to \left\{ {\left( {{\cal O}\backslash \left\{ {{{\cal G}_a},{{\cal G}_b}} \right\}} \right) \cup \left( {{{\cal G}_a}\backslash \left\{ n \right\}} \right) \cup \left( {{{\cal G}_b} \cup \left\{ n \right\}} \right)} \right\}$.
\end{definition}

The coalition-friendly preference rule is considered from a coalition standpoint, which can be viewed as a partially collaborative approach. As a result, it is critical to investigate the stability under it.

\begin{definition}
If there exists a potential function $\phi $ such that the difference between the potential function and the utility function remains constant when the client's association relationship changes, the game is an exact potential game.
    \begin{equation}
        \phi \left ( \widetilde{a_n} ,a_{-n} \right ) -\phi \left ( a_n ,a_{-n} \right ) = {\cal U}_n(\widetilde{a_n} ,a_{-n})-{\cal U}_n(a_n ,a_{-n}).
    \end{equation}
\end{definition}
\begin{theorem}
    The coalition formation game $\mathcal{C}$ is an exact potential game.
    \label{EPG}
\end{theorem}

\begin{algorithm}[tb]
    \caption{Coalition Formation Game for Data Distributions Adjustment}
    \label{alg:algorithm1}
  \KwIn{${\cal N} = \left\{ {1,\cdots, N} \right\}$, ${\cal M} = \left\{ {1, \cdots, M} \right\}$, ${{\cal O}_{cr}}$, and $L^{max}$}
  \KwOut{Final partition ${{\cal O}^*} = \left\{ {{\cal G}^*} \right\}_1^M$}
  ${{\cal O}^*} = \emptyset$, $l = 0$\;
  \Repeat{coalition partition converges or $l = L^{max}$}{
	$n = random\left\{1, \cdots, N\right\}$, $n\in {\cal{G}}_m$\;
	\ForEach{${{\cal G}_{m'}} \in {{\cal O}_{cr}}, m \ne m'$}
	{
		Calculate $ \mathcal{\overline{JS}}_{\mathcal{G}_m \to \mathcal{G}_{{m}'}}^n$\;
	}
	${m}' =\underset{m}{\min} \left \{  \mathcal{\overline{JS}}_{\mathcal{G}_m \to\mathcal{G}_1}^n, \cdots, \mathcal{\overline{JS}}_{\mathcal{G}_m \to\mathcal{G}_M}^n\right \} $\;
	\If{${m}' != m$}
	{
		${{\cal G}_m} = {{\cal G}_m}\backslash \left\{ n \right\}$, ${{\cal G}_{m'}} = {{\cal G}_{m'}} \cup \left\{ n \right\}$\;
		${{\cal O}_{cr}} = \left( {{{\cal O}_{cr}}\backslash \left\{ {{{\cal G}_m},{{\cal G}_{m'}}} \right\}} \right) \cup \left( {{{\cal G}_m}\backslash \left\{ n \right\}} \right) \cup \left( {{{\cal G}_{m'}} \cup \left\{ n \right\}} \right)$\;
	}
	 $l = l+1$\;
}
\end{algorithm}

According to Theroem \ref{EPG}, the coalition formation game $\cal C$ has at least a stable coalition partition. To obtain the solution of the game, we will focus on the algorithm for forming an effective coalition partition, which shown in Algorithm \ref{alg:algorithm1}. In the coalition formation algorithm consisting of $N$ clients and $M$ ESs, the initial coalition partition ${{\cal O}_{cr}}$ is first formed by randomly associating clients with ESs. Then, a client $n$ is selected to undergo a comparative update based on the switch rule defined in Definition \ref{定义_Switch rule}. This rule determines whether the client should leave its current coalition or join another coalition (line 3-7). We assume that client $n$ leave current coalition and compute $ \mathcal{\overline{JS}}$ of each situation that client $n$ join in other coalition respectively based on Eq. (\ref{JS散度计算_Avg}). Therefore, according to the result of assumption, the prioritization of each situation or coalition can be determined. We choose the case that yields the lowest $ \mathcal{\overline{JS}}$, and then client $n$ leaves the current coalition ${{\cal G}_m}$, joins the new coalition ${{\cal G}_{m'}}$ if the two coalition are not the same (line 8-9). Coalition partition will be updated due to this switching (line 10). However, if $ \mathcal{\overline{JS}}$ increases after the switch operation, client will remain in the current coalition. The iterative process described above repeats until form a stable partition of coalitions ${{\cal O}^*} = \left\{ {{\cal G}^*} \right\}_1^M$ where no exchange exists that can bring down $ \mathcal{\overline{JS}}$ in current partition ${{\cal O}_{cr}}$ or reach the maximum iteration rounds. We need to perform $\frac{\left ( m-1 \right ) m}{2}$ calculations for $ \mathcal{\overline{JS}}$. However, in reality, the computation is equally distributed to each ES, so the final time complexity is $O\left ( M\right ) $. This is hardly a burden for a high-performance ES that can respond quickly to clients.

\subsection{Optimal Bandwidth Allocation}
Based on the final coalition partition, the delay of local training is determined. According to the definitions of energy consumption of local computing and transmission from Eq. (\ref{能耗_计算}) to Eq. (\ref{能耗_总}), energy consumption is proportional to the execution time of training. From Eq. (\ref{时间_通信}) and Eq. (\ref{eq:constraint2}), we can observe that the transmission delay is minimized when $p_n^* = p_n^{max} $,$\forall n \in {\cal N}$ . Hence, the joint optimization problem Eq. (\ref{eq:Max&Min}) is distilled into a single-object optimization problem, as expressed in Eq. (\ref{eq:Min}).

\begin{subequations}
\begin{align}
{\mathbb P_2 }: \quad &  \underset{\mathbf{B}}{\min} \enspace {\cal T}^{\cal U} , \label{eq:Min}\\
& \text{s.t.} \quad \sum_{m=1}^{M} B_m=B, \forall m\in \mathcal{M}, \label{eq:Min_constraint2} \\
& \quad \quad  B_m > 0, \forall m\in \mathcal{M} .\label{eq:Min_constraint3} 
\end{align}
\end{subequations}

It can be observed that ${\cal T}_n^{\cal U}(B_m)$ is a convex function with respect to $B_m$ from Eq. (\ref{时间_通信}). We assume that the worst transmission case in each coalition is $n_m^0$ and the clients in each coalition have the same status as the worst case. The solution of the communication minimization problem ${\mathbb{P}_2}$, i.e., optimal bandwidth allocation for coalitions, is denoted as $\mathbf{B}^*$. Then, the $\lambda _2 E$ can reach the minimum value when $\mathbf{B}=\mathbf{B}^*$. Because a strictly convex function has at most one minimum, by setting $n_m = n_m^0$, $\forall m\in{\cal M}$ and $p_{n,m}^* = p_{n,m}^{max} $, $\forall n \in {\cal N}$, the optimization problem is transformed as:

\begin{subequations}
\begin{align}
{\mathbb{P}_3 }:& \quad \underset{\mathbf{B} }{\min} \enspace \sum_{m=1}^{M} \lambda_2 \left | \mathcal{G}_m  \right |  \tau_g \tau_e  \frac{p_{n_m^0}^{max}\mathbb{Z}}{\frac{B_m}{\left | \mathcal{G}_m  \right |}log_2\left ( 1+\frac{p_{n_m^0}^{max}h_{n,m}}{\frac{B_m}{\left | \mathcal{G}_m  \right |}\mathbb{N}_0}  \right )  } , \\
& \quad \text{s.t.} \quad  \sum_{m=1}^{M} B_m=B, \forall m\in \mathcal{M} ,\label{eq:Min_constraint2} \\
& \quad \quad \quad  B_m > 0, \forall m\in \mathcal{M} .\label{eq:Min_constraint3} 
\end{align}
\end{subequations}

\begin{lemma}
    If $g_i(x)$ is convex function, $\max(\min)g_i(x)$, $\sum g_i(b_ix)$ and $\sum b_i g_i(x)$ are also convex functions.
    \label{凸性}
\end{lemma}

Since the objective function of optimization problem is the sum of a convex function, according to Lemma \ref{凸性}, the objective function is also convex with respect to $\mathbf{B}$.

We can apply the gradient projection method (GP) \cite{GP} to allocate bandwidth when coalition partition is determined. The GP method is summarized in Algorithm 2. Through several iterations from line 3 to line 6, we can obtain the optimal bandwidth allocation of the problem specified in ${\mathbb{P}_2}$ and ${\mathbb{P}_3}$.

\subsection{Optimal Transmit Power}
Once the bandwidth allocation matrix $\mathbf{B}^*$ and stable coalition partition $\mathcal{G}^*$ (or edge association matrix $\mathbf{A}^*$) are determined, the optimization of transmit power for each client can be formulated as follows:
\begin{subequations}
\begin{align}
{\mathbb{P}_4 }:  \quad & \underset{\mathbf{p}}{\min} \enspace \lambda _2 {\cal E}^{\cal U} , \label{eq:Min_E}\\
& \text{s.t.} \quad p_n\in \left ( 0,p_n^{max} \right ], \forall n\in \mathcal{N}, \label{eq:Min_E_constraint4} \\
& \quad \quad {\cal T}_{n,t}\le \frac{\mathbb{I}}{\tau_e\tau_g}, \forall n\in \mathcal{N}. \label{eq:Min_E_constraint5}
\end{align}
\end{subequations}
Eq. (\ref{eq:Min_E_constraint4}) gives the clients' transmission power range and Eq. (\ref{eq:Min_E_constraint5}) emphasizes the constraint of maximum execution latency, so the solution needs to satisfy both of them.

\begin{theorem}
\label{theorem2}
There exists an optimal solution $p_{n,m}^*$ for problem $\mathbb{P}_4$, i.e.,

\begin{equation}
    p_{n,m}^*=\min\left \{{p_n^{max}} ,p_{n,\mathbb{I} }\right \}  ,
\end{equation}
where
\begin{equation}
    p_{n,\mathbb{I} }=\frac{B_{n,m}^U\mathbb{N}_0\left ( 2^{\frac{\mathbb{Z}}{B_{n,m}^U\left ( \frac{\mathbb{I}}{\tau_e\tau_g}-\tau_c {\cal T}_{n,t}^C  \right ) } }-1 \right ) }{h_{n,m}} .
\end{equation}
\end{theorem}

\section{Experiments}
In this section, we conduct extensive experiments to assess the performance of LEAP. We first descript the experimental environments, and then compare and analyze the effectiveness of our scheme in comparison to other methods.

\subsection{Experimental Setup}

\textbf{Datasets and Models.} We evaluate the performances of LEAP on two commonly adopted learning models and four real datasets: a LR (Logistic Regression) model on MNIST datatset \cite{MNIST} and a CNN (convolutional neural network) with two convolution layers and three fully connected layers on CIFAR-10 \cite{Cifar}, SVHN \cite{SVHN} and CINIC-10 \cite{CINIC10}.

\begin{algorithm}[tb]
    \caption{Bandwidth Allocation}
    \label{alg:algorithm}
  \KwIn{ $\mathbf{B}(0)$, step size $\eta$, accuracy tolerance $\epsilon$, and iteration number $J^{max}$}
  $j = 0$\;
  \Repeat{objective value converges or $ j= J^{max}$}{
	Gradient: $\nabla G\left ( \mathbf{B}_j \right ) $\;
	Projection: $P_{\Omega _\mathbf{B}}$\;
 	Update $\mathbf{B}$: $\mathbf{B}_{j+1}\longleftarrow P_{\Omega_\mathbf{B} }\left ( \mathbf{B}_j- \eta \nabla G\left ( \mathbf{B}_j \right )  \right ) $\;
	$j=j+1$\;
}
\end{algorithm}

\begin{figure*}[htbp]
\begin{center}
\subfigure[$\overline{\mathcal{JS}}=0.69$]{
	\begin{minipage}[b]{.235\linewidth}
        \centering
        \includegraphics[scale=0.12]{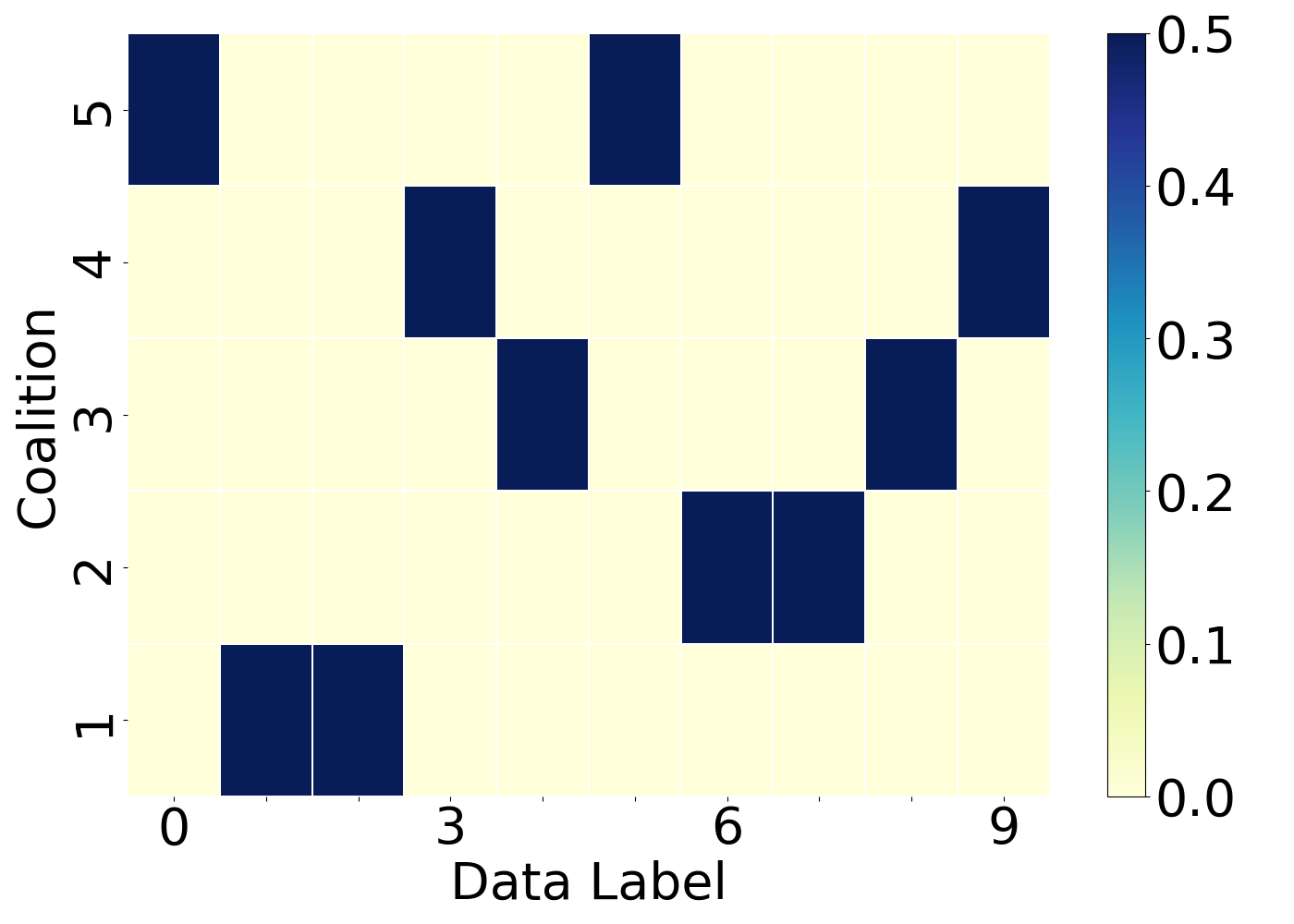}
\label{AvgJSD_1}
 	\vspace{-1em}
    \end{minipage}
}
\subfigure[$\overline{\mathcal{JS}}=0.49$]{
	\begin{minipage}[b]{.235\linewidth}
        \centering
        \includegraphics[scale=0.12]{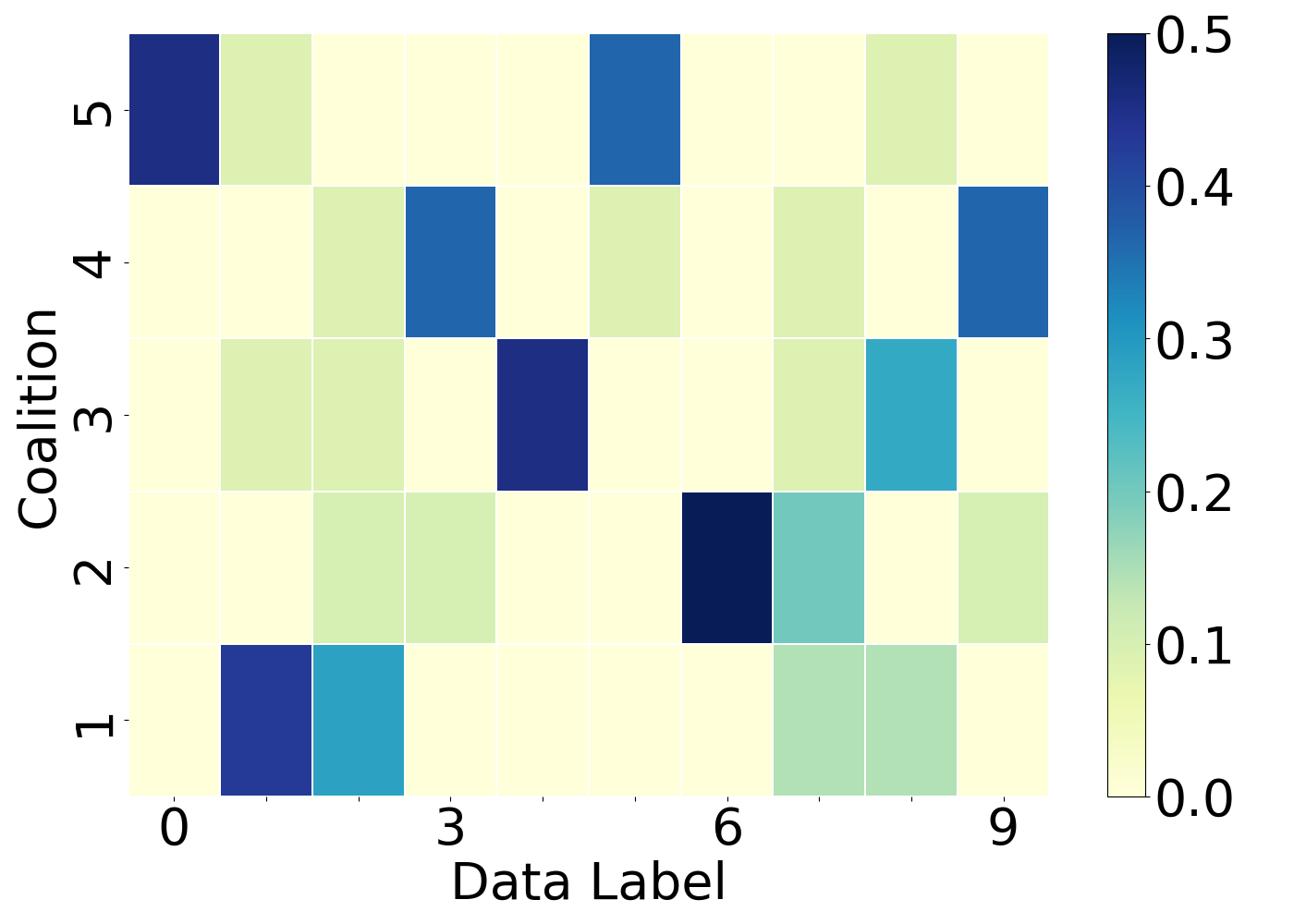}
\label{AvgJSD_2}
 	\vspace{-1em}
    \end{minipage}
}
\subfigure[$\overline{\mathcal{JS}}=0.0$]{
	\begin{minipage}[b]{.235\linewidth}
        \centering
        \includegraphics[scale=0.12]{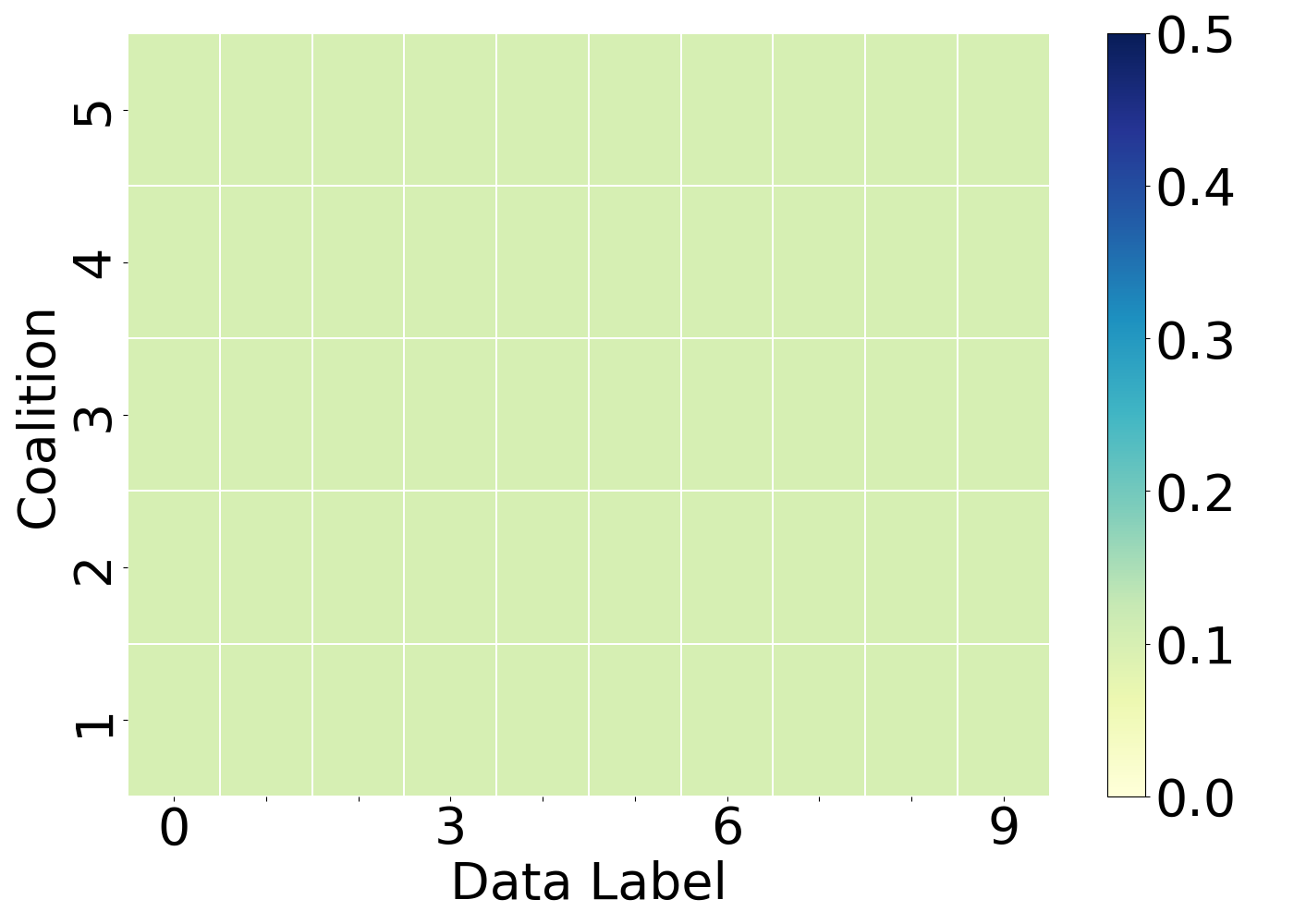}
	 \vspace{-1em}
\label{AvgJSD_3}
    \end{minipage}
}
\subfigure[$\overline{\mathcal{JS}}$'s change process]{
	\begin{minipage}[b]{.235\linewidth}
        \centering
        \includegraphics[scale=0.12]{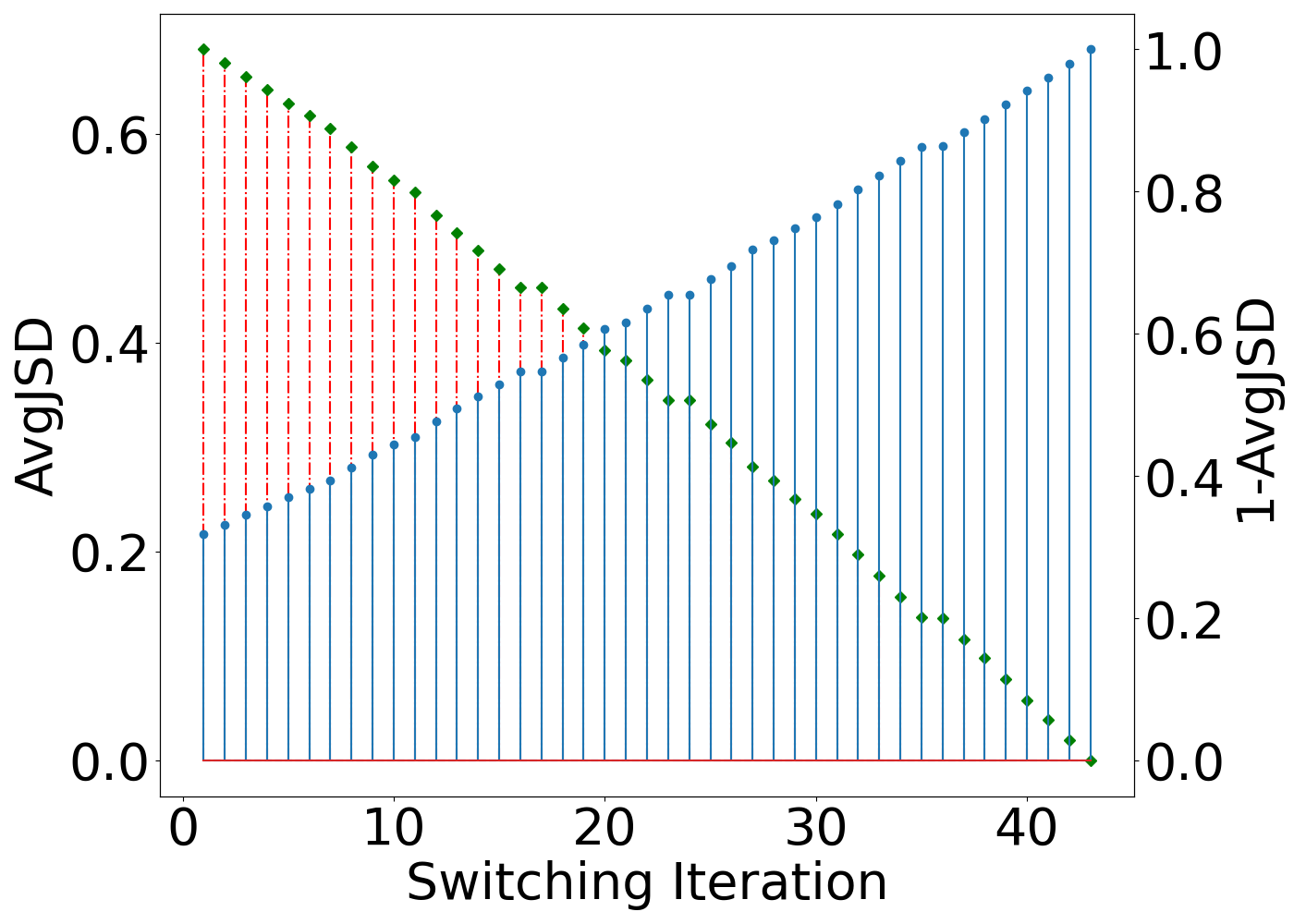}
	 \vspace{-1em}
\label{AvgJSD_change}
    \end{minipage}
}
 \vspace{-1em}
\caption{Changes of data distribution and $\mathcal{\overline{JS}}$ during coalition formation.}

 \vspace{-0.6em}
\end{center}
 \vspace{-0.6em}
\end{figure*}

\begin{figure*}[t]
\begin{center}

\subfigure[MNIST]{
	\begin{minipage}[b]{.23\linewidth}
        \centering
        \includegraphics[scale=0.12]{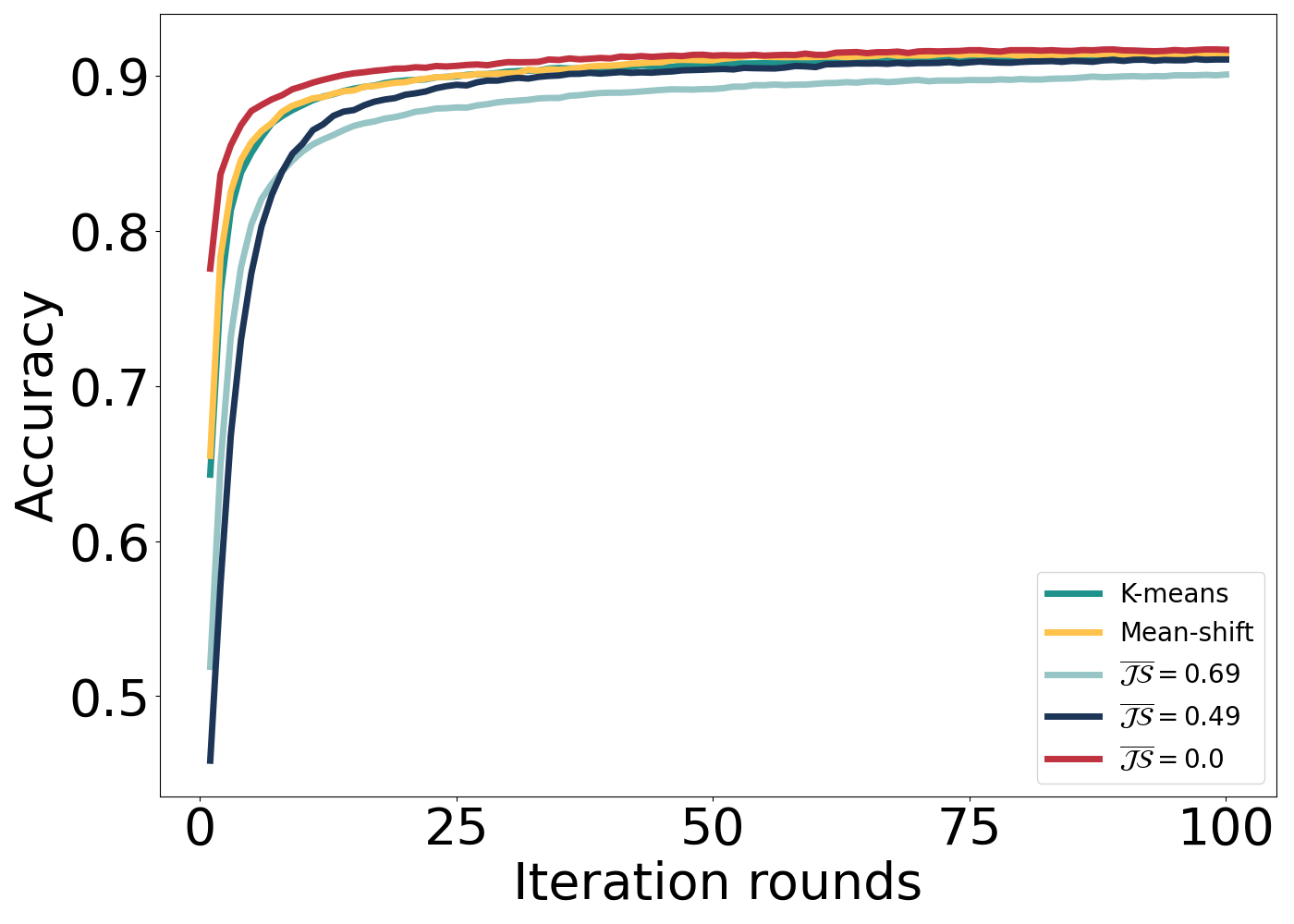}
 \vspace{-1em}
    \end{minipage}
\label{mnist}
}
\subfigure[CIFAR-10]{
	\begin{minipage}[b]{.23\linewidth}
        \centering
        \includegraphics[scale=0.12]{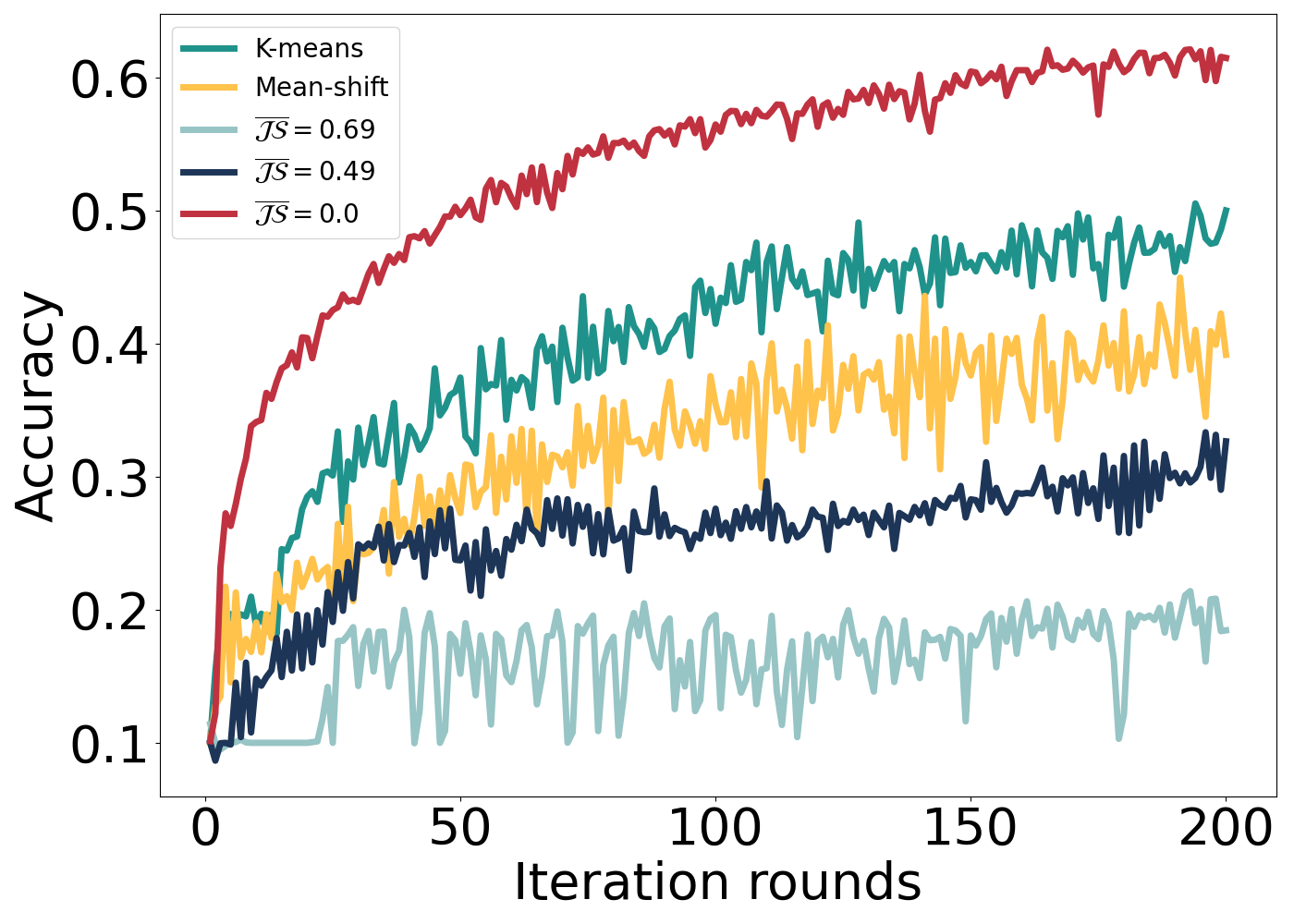}
 \vspace{-1em}
    \end{minipage}
\label{cifar10}
}
\subfigure[SVHN]{
	\begin{minipage}[b]{.23\linewidth}
        \centering
        \includegraphics[scale=0.12]{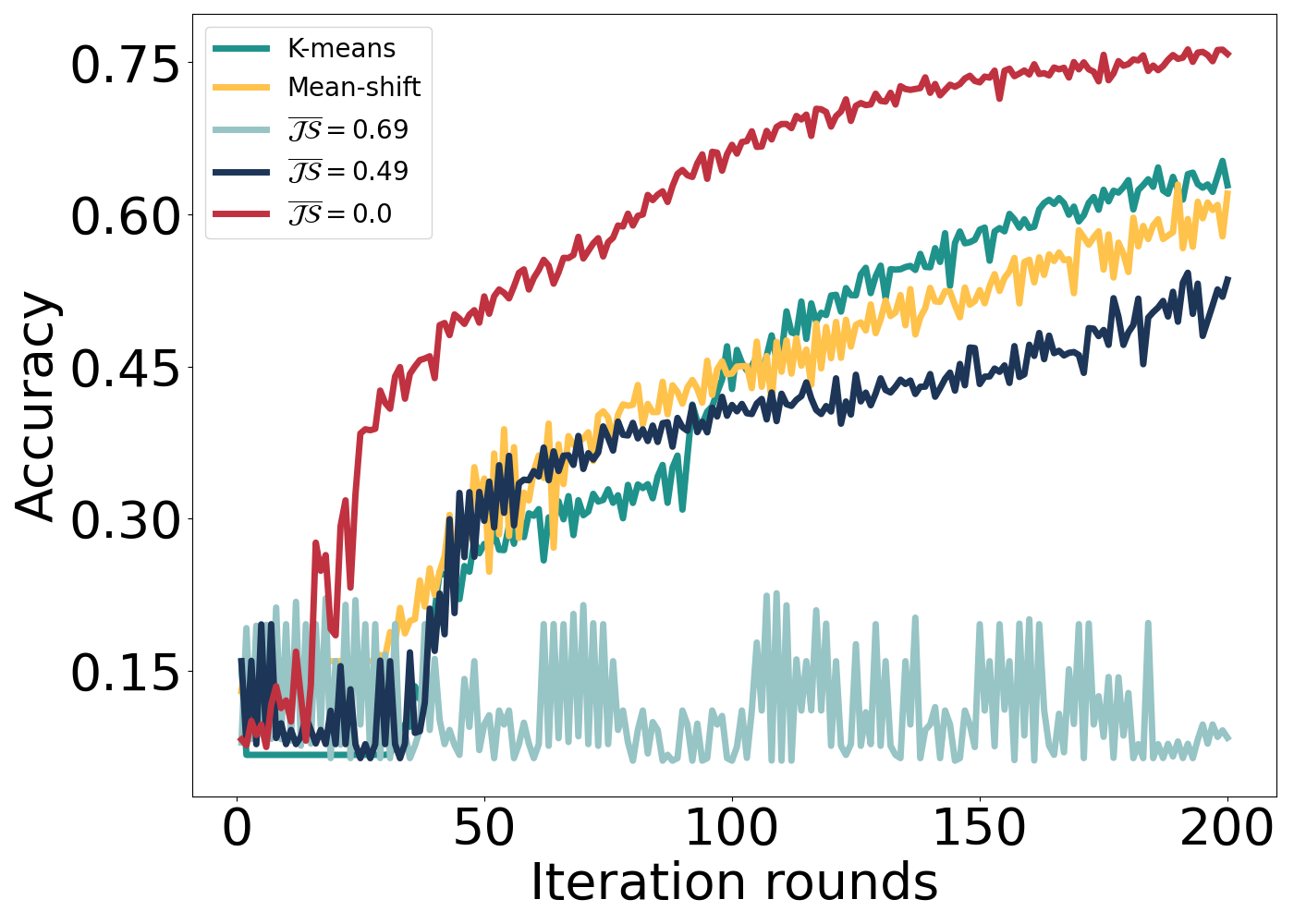}
 \vspace{-1em}
    \end{minipage}
\label{svhn}
}
\subfigure[CINIC-10]{
	\begin{minipage}[b]{.23\linewidth}
        \centering
        \includegraphics[scale=0.12]{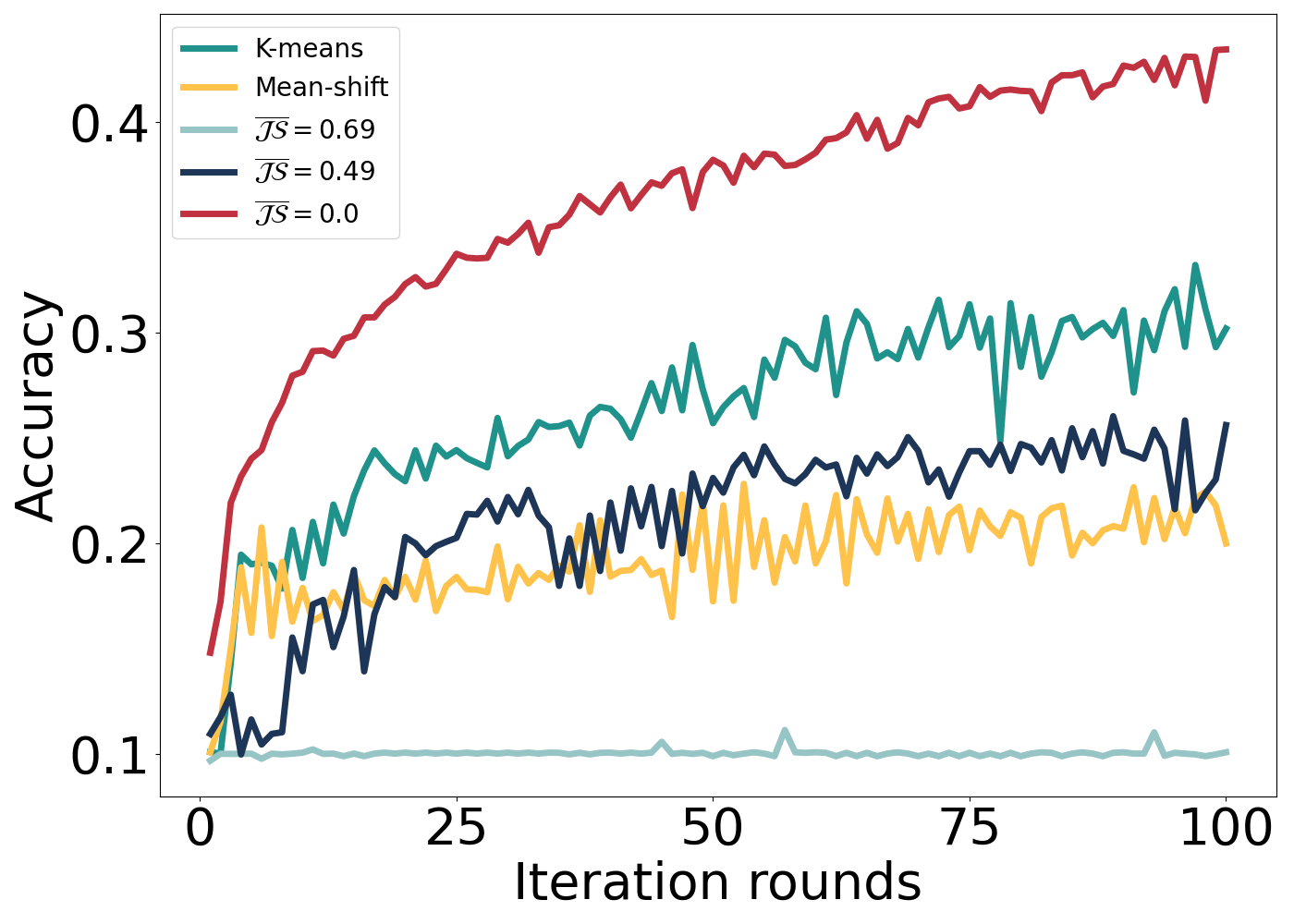}
 \vspace{-1em}
    \end{minipage}
\label{cinic10}
}
 \vspace{-1em}
\caption{Global model performance comparison of different data distributions and methods on four datasets.}
\label{333}
 \vspace{-0.6em}
\end{center}
 \vspace{-0.6em}
\end{figure*}

\textbf{Parameter Settings.} We set two scale settings on each dataset, 5 ESs with 50 clients or 3 ESs with 10 clients. For each dataset, 5 rounds of local training and 12 rounds of edge iterations are conducted. Among these setting, 100 rounds of global iterations are performed for MNIST and CINIC-10 datasets, while 200 rounds of global iterations for CIFAR-10 and SVHN. We set learning rate to 0.01 for MNIST, CIFAR10 and CINIC-10, and 0.005 for SVHN. The momentum is set within the range of $\left[0,0.9\right]$ and weight decay is 0.005.

\textbf{Baselines.} Four baselines are considered for comparison with LEAP, consist of two clusting alorithms, a coalition formation method and a model aggregation method.
\begin{itemize}
	\item \textbf{Mean-shift} \cite{meanshift}: Mean-Shift algorithm is a density-based non-parametric clustering algorithm. One key advantage is that it does not require specifying the number of clusters in advance, as it automatically determines it based on the data distribution.

	\item \textbf{K-means} \cite{kmeans}: K-Means algorithm is an iterative clustering algorithm that partitions data points into K clusters (K is pre-specified) and assigns each data point to the nearest cluster center based on distance.

    \item \textbf{RH} \cite{HFL_step}: RH is a reputation-aware hedonic coalition formation algorithm, in which clients form stable coalition partitions with selfish preferences based on the reputation of cluster heads and their own utility.

    \item  \textbf{MA} \cite{MA1} \cite{MA}: MA is a model aggregation method based on marginal losses. By setting marginal loss thresholds, it becomes possible to identify and exclude low-quality models or reduce their contribution to the aggregation process.

\end{itemize}

\subsection{Experimental Results} \textbf{Validating the effectiveness of mitigating the degree of cross-edge non-IID.} \cref{AvgJSD_1,AvgJSD_2,AvgJSD_3} show the distribution of data for each coalition during the coalition formation process, with the color of each cell indicating the percentage of such data under that coalition. The initial $\overline{\mathcal{JS}}$ is 0.69 with two label categories of each coalition. As the client switching process progresses, the data distributions under each colaition become increasingly similar, with the final $\overline{\mathcal{JS}}$ reaching 0, which means that the distribution in each coalition is same. Moreover, the color of all cells is similar, indicating that all data categories are equally represented in each coalition. \cref{AvgJSD_change} shows the complete variation of $\overline{\mathcal{JS}}$ during the client switching process. Each switching operation demonstrates a consistent decreasing trend of $\overline{\mathcal{JS}}$.

\textit{Comparing with K-means and Mean-shift algorithms.} Fig. \ref{333} shows the accuracy under different methods and different data distributions. Comparing the initial state, the average accuracy based on the final distribution is improved by 2.9\%, 33.3\%, 47.6\%, and 26.2\% in the four datasets, respectively. Based on the same initial conditions, Mean-shift algorithm divides the data into five clusters. While randomly assigning the clients to ESs, the optimal client combination cannot be ensured because of the duplicate labels within the clusters. A similar issue arises when using the K-means algorithm. In addition, the K-means algorithm requires specifying the number of clusters in advance, which further hampers its applicability. It is clear that the final result after optimization based on our method is significantly improved compared to the other two methods, because our method is always in the direction of better when adjusting the combination of data distribution.

\begin{table}[t]
\caption{Average model accuracy of different methods based on the finial coalition partition in RH.}
\vspace{-0.6em}
\setlength{\tabcolsep}{1pt}
\begin{tabularx}{0.48\textwidth}{XXXXX}
\toprule
\multirow{2}{*}{\centering Methods} & \multicolumn{4}{c}{Datasets} \\ 
\cmidrule{2-5}
                          & MNIST      & CIFAR-10  & SVHN     & CINIC-10 \\ 
\midrule
RH                        & 88.65\%    & 51.19\%   & 68.93\%   & 32.95\%  \\ 
MA                        & 80.00\%    & 36.31\%   & 53.07\%   & 24.29\%   \\ 
Our                       & \textbf{90.75\%}    & \textbf{58.63\%}   & \textbf{73.77\%}   & \textbf{40.14\%}  \\ 
\bottomrule
\end{tabularx}
\vspace{-1em}
\label{table1}
\end{table}

\textit{Comparing with RH and MA.} The the initial correlation relation of Table \ref{table1} is based on the experimental results in \cite{HFL_step} that presented RH, with 10 clients and 3 ESs. Compared to RH and MA, our approach still performs well. This is because RA performs association formation with selfish client preference rule without considering the impact on the coalition partition. MA discards some of the model parameters below the loss threshold when aggregating based on marginal losses, resulting in data wasting.

\begin{figure}[tbp]
\vspace{-0.6em}
\begin{center}
\subfigure[Transmission Energy]{
	\begin{minipage}[t]{0.46\linewidth}
\centering
\includegraphics[scale=0.122]{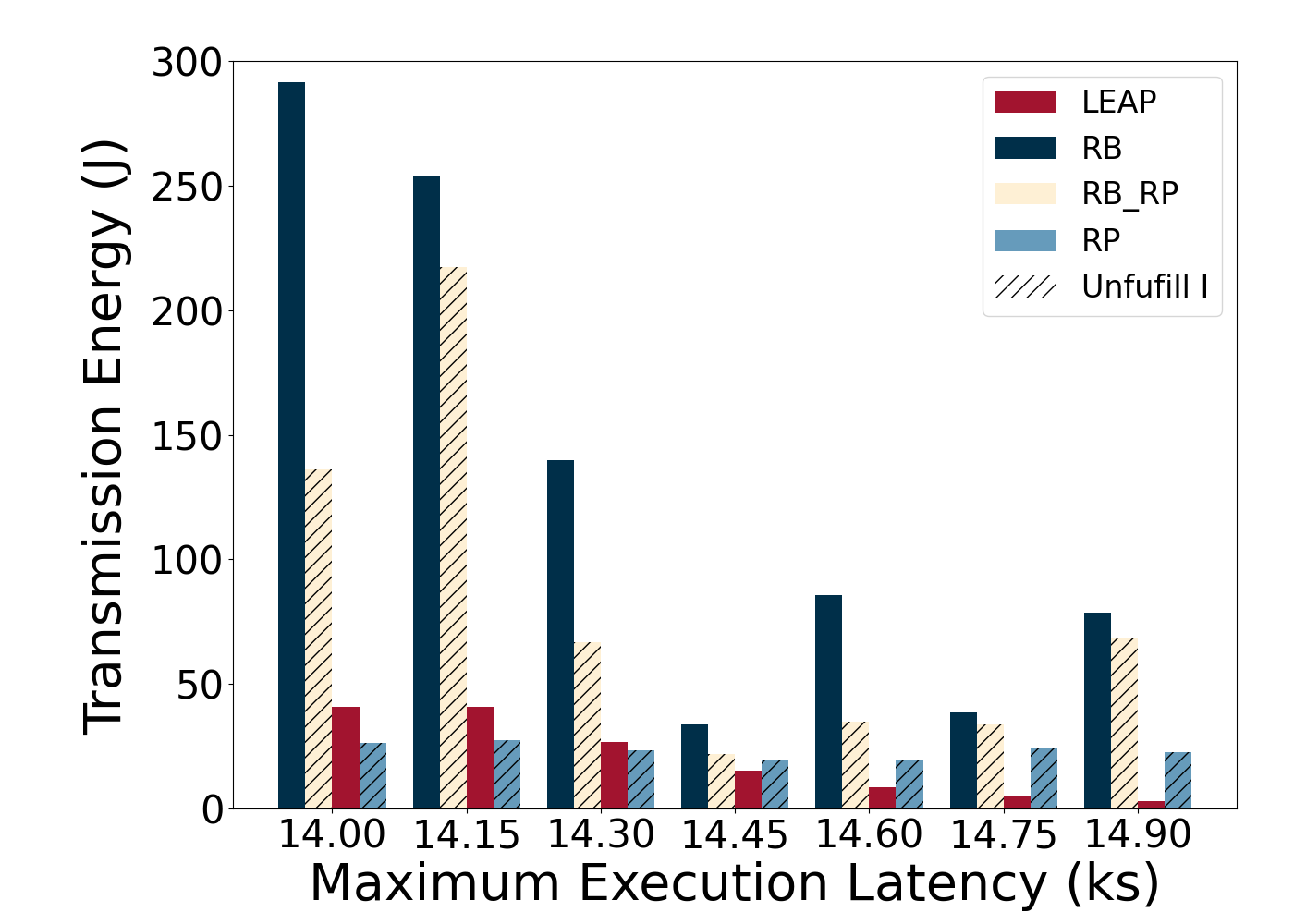}
\end{minipage}
\label{TE}
}
\subfigure[Transmission Power]{
	\begin{minipage}[t]{0.46\linewidth}
\centering
\includegraphics[scale=0.122]{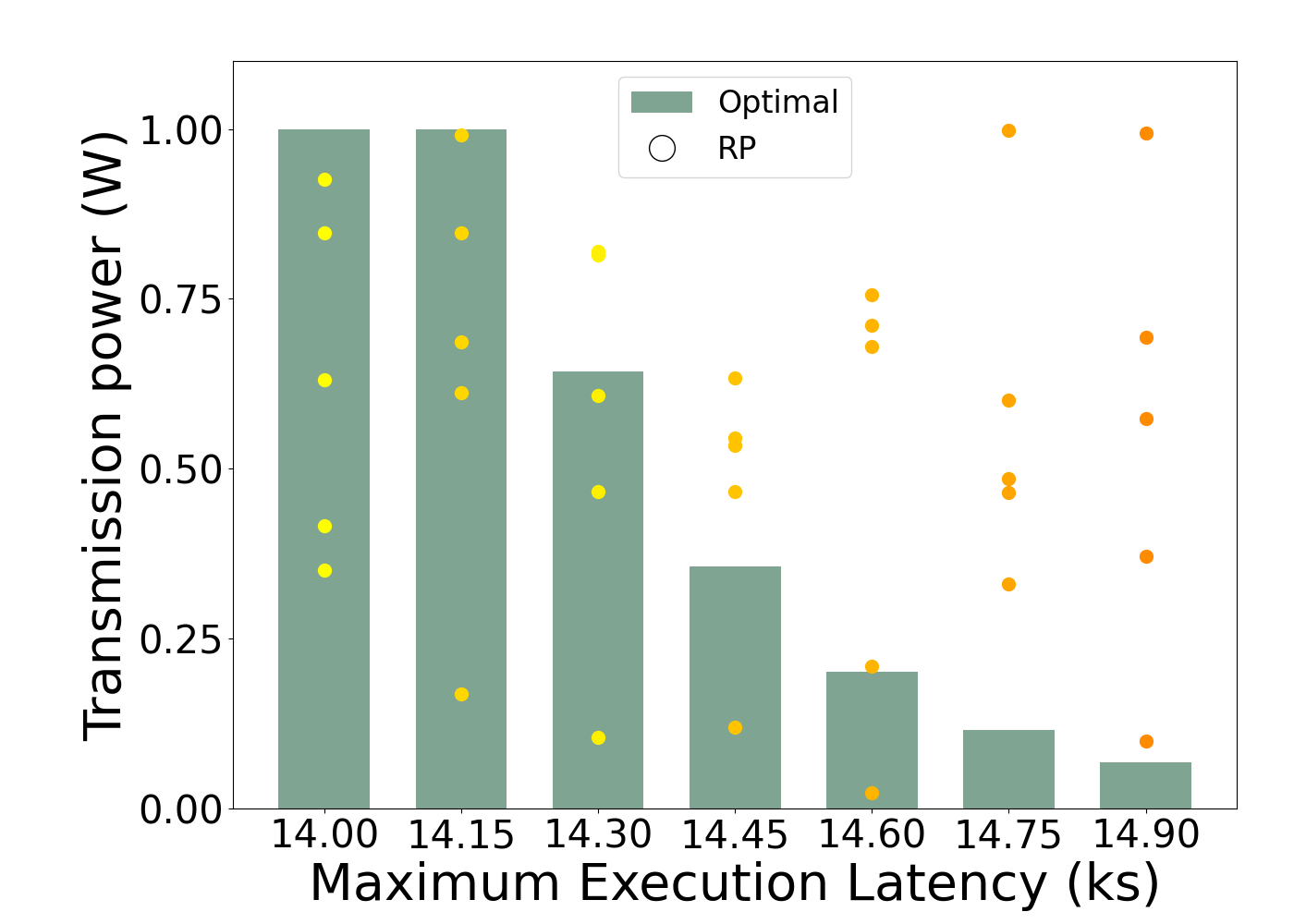}
\end{minipage}
\label{TP}
}
\vspace{-1em}
\caption{Transmission energy consumption and transmission power under different optimization schemes.}
\end{center}
\vspace{-2em}
\end{figure}

\textbf{Verifing the effectiveness on resource allocation.} We calculate average transmission energy consumed per round of edge aggregation with random bandwidth allocation (RB), random transmission power (RP), and a combination of RB and RP (RB\_RP). From Fig. \ref{TE}, we can observe that LEAP achieves a significant reduction in transmission energy consumption. We notice that in some cases RP be lower, but it fails to satisfy the maximum execution delay. From Fig. \ref{TP}, the randomly determined transmission power is below the optimal value several times, so it fails to satisfy the delay requirement despite producing lower energy consumption.

\section{Conclusion}
In this article, a novel optimization method LEAP, which has a lightweight implementation, was proposed to address the impact of muti-dimentional properties on HFL, i.e., data contribution, consumption of time and energy. Due to the stochastic nature of data distribution under ESs, edge association was combined with LEAP and a coalition formation game was built to model data distribution under different associations. To reduce the degree of non-IID of cross-edge data, coalition-friendly preference rule was employed, and the existence of stable coalition partitions was proved. Further, the gradient projection method (GP) was utilized to reduce task execution time in heterogeneous resources within stable coalitions, improving the communication efficiency. Finally, extensive experiments were conducted on various real datasets to validated the effectiveness of LEAP.

\newpage
\section*{Acknowledgments}
This work was supported in part by the National Natural Science Foundation of China (No. 62372343, 62072411), in part by the Zhejiang Provincial Natural Science Foundation of China (No. LR21F020001), and in part by the Key Research and Development Program of Hubei Province (No. 2023BEB024).

\appendix
\section*{Appendix}
\subsection*{Proof of Theorem 1}
The potential function $\phi$ of the coalition formation game is defined as the sum of $\overline{\mathcal{JS}}$, calculated as follows:
    \begin{equation}
        \phi \left ( {a_n} ,a_{-n} \right ) = \sum_{m=1}^{M} \overline{\mathcal{JS}}\left ( Q_{a_n},Q_{a_{-n}} \right ) .
    \end{equation}
When client $n$'s association relationship changes from $a_n$ to $\widetilde{a_n}$, the corresponding change in the potential function is
    \begin{equation}
    \begin{aligned}
         &  \phi \left ( \widetilde{a_n} ,a_{-n} \right ) -\phi \left ( a_n ,a_{-n} \right ) \\
        & =  \sum_{\underset{n \in \mathcal{G} _{\widetilde{a_n}}}{i=1,} }^{M-1} \sum_{j=i+1}^{M} \mathcal{JS}(Q_i,Q_j)  - \sum_{\underset{n \in \mathcal{G} _{a_n}}{i=1,} }^{M-1} \sum_{j=i+1}^{M} \mathcal{JS}(Q_i,Q_j)\\
        & = \sum_{j=\widetilde{a_n}+1 }^{M} \left [ \mathcal{JS}(Q_{\widetilde{a_n}\cup\left \{ n\right \}},Q_j)-\mathcal{JS}(Q_{\widetilde{a_n}\setminus \left \{ n\right \}},Q_j) \right]\\
        & \quad + \sum_{j=a_n+1 }^{M} \left [ \mathcal{JS}(Q_{a_n\setminus \left \{ n\right \}},Q_j)-\mathcal{JS}(Q_{a_n\cup  \left \{ n\right \}},Q_j) \right ]  \\
	   & \quad +\sum_{i=1}^{\widetilde{a_n}-1}\left[\mathcal{JS}(Q_i,Q_{\widetilde{a_n}\cup\left \{ n\right \}}) -\mathcal{JS}(Q_i,Q_{\widetilde{a_n}\setminus\left \{ n\right \}}) \right] \\
	   & \quad + \sum_{i=1}^{{a_n}-1}\left[ \mathcal{JS}(Q_i,Q_{{a_n}\setminus \left \{ n\right \}})- \mathcal{JS}(Q_i,Q_{{a_n}\cup \left \{ n\right \}}) \right]\\
        & \quad +\sum_{i\in \Omega}^{M-1} \sum_{\underset{j\in \Omega}{j=i+1,} }^{M} \left [ \mathcal{JS}(Q_i,Q_j)-\mathcal{JS}(Q_i,Q_j) \right ] ,
    \end{aligned}
    \end{equation}
    where $\Omega  $ represents the rest of ESs or coalitions in $\left \{ \mathcal{M}\setminus \left \{ a_n \right \} \setminus \left \{ \widetilde{a_n} \right \}   \right \} $. According to Lemma \ref{凸性}, we know that other coalitions in $\Omega  $  will be unaffected by client $n$’s changes. The JSD value of the coalition joined by client $n$ can be expressed as 
    \begin{equation}
    \begin{aligned}
		& U_n(a_n ,a_{-n})\\
		&  =\sum_{j=a_n+1}^{M-1} \mathcal{JS}(Q_{a_n},Q_j) + \sum_{j=\widetilde{a_n} +1}^{M-1} JS(Q_{\widetilde{a_n} },Q_j) \\
		& \quad +\sum_{i=1}^{{a_n}-1}\mathcal{JS}(Q_i,Q_{{a_n}}) + \sum_{i=1}^{\widetilde{a_n}-1}\mathcal{JS}(Q_i,Q_{\widetilde{a_n}}) \\
		& \quad +\sum_{i\in \Omega  }^{M-1} \sum_{\underset{j\in \Omega}{j=i+1,}}^{M}\mathcal{JS}(Q_i,Q_j) .
    \end{aligned}
    \end{equation}
And the difference in JSD value can be described as follows:
    \begin{equation}
    \begin{aligned}
         & U_n(\widetilde{a_n} ,a_{-n})-U_n(a_n ,a_{-n}) \\
         &=\sum_{j=\widetilde{a_n}  +1}^{M-1} \mathcal{JS}(Q_{\widetilde{a_n}\cup \left \{ n \right \}  },Q_j)+\sum_{j=a_n+1}^{M-1} \mathcal{JS}(Q_{a_n\setminus \left \{ n \right \} },Q_j)\\
         & \quad +\sum_{i=1}^{\widetilde{a_n}-1}\mathcal{JS}(Q_i,Q_{\widetilde{a_n}\cup\left \{ n\right \}}) + \sum_{i=1}^{{a_n}-1}\mathcal{JS}(Q_i,Q_{a_n\setminus \left \{ n \right \}})\\
         & \quad -\sum_{j=a_n+1}^{M-1} \mathcal{JS}(Q_{a_n\cup  \left \{ n \right \} },Q_j) - \sum_{j=\widetilde{a_n}+1}^{M-1} \mathcal{JS}(Q_{\widetilde{a_n}\setminus \left \{ n \right \} },Q_j)\\
	    & \quad -\sum_{i=1}^{{a_n}-1}\mathcal{JS}(Q_i,Q_{{a_n}\cup\left \{ n\right \}}) - \sum_{i=1}^{\widetilde{a_n}-1}\mathcal{JS}(Q_i,Q_{\widetilde{a_n}\setminus \left \{ n \right \}})\\
         & \quad + \sum_{i\in \Omega}^{M-1}\sum_{\underset{j\in \Omega}{j=i+1,} }^{M-1} \mathcal{JS}(Q_{i},Q_j)-\sum_{i\in \Omega  }^{M-1} \sum_{\underset{j\in \Omega}{j=i+1,} }^{M}\mathcal{JS}(Q_i,Q_j)\\
         & = \phi \left ( \widetilde{a_n} ,a_{-n} \right ) -\phi \left ( a_n ,a_{-n} \right ) .
    \end{aligned}
    \end{equation}
    Since, the coalition formation game, as an exact potential game, has at least one pure strategy Nash equilibrium, Theroem \ref{EPG} is proved.

\subsection*{Proof of Theorem 2}
The objective function in problem $\mathbb{P}_4$ can be expressed as
\begin{equation}
	\begin{aligned}
		\quad \lambda_2 {\cal E}^{\cal U} = \lambda_2 \sum_{n \in {\cal G}_m}  \tau_g \tau_e {\cal E}_{n,t}^{\cal U}.
	\end{aligned}
\label{eq28}
\end{equation}
By analyzing the function of
\begin{equation}
	{\cal E}_{n,t}^{\cal U}=\frac{\mathbb{Z} p_{n,m}}{B_{n,m}^U\log_{2}\left ( {1+\frac{p_{n,m}h_{n,m} }{B_{n,m}^U \mathbb{N}_0}}   \right )},
\end{equation}
the first derivative of $\mathcal{E}_{n,m}^{U}$ with respect to $p_{n,m}$ is
\begin{equation}
\frac{\partial {\cal E}_{n,t}^{\cal U}}{\partial p_{n,m}}=\frac{\mathbb{Z}}{B_{n,m}} \frac{\log_{2}\left ( {1+\frac{p_{n,m}h_{n,m} }{B_{n,m}^U \mathbb{N}_0}}\right )-\frac{p_{n,m}h_{n,m}}{\left ( B_{n,m}^U \mathbb{N}_0 + p_{n,m}h_{n,m}\right )\ln{2}  } }{\log_{2}^2\left ( {1+\frac{p_{n,m}h_{n,m} }{B_{n,m}^U \mathbb{N}_0}}\right )} .
\end{equation}
By setting $\varsigma = {1+\frac{p_{n,m}h_{n,m} }{B_{n,m}^U \mathbb{N}_0}}> 1$, we can derive that

\begin{equation}
	\begin{aligned}
		&  \log_{2}\left ( {1+\frac{p_{n,m}h_{n,m} }{B_{n,m}^U \mathbb{N}_0}}\right )-\frac{p_{n,m}h_{n,m}}{\left ( B_{n,m}^U \mathbb{N}_0 + p_{n,m}h_{n,m}\right )\ln{2}  } \\
	& = \log_{2}\left ( \varsigma \right ) - \frac{1-{\frac{1}{ \varsigma}}}{\ln{2}} > 0.
	\end{aligned}
\end{equation}
Therefore, we know that $\frac{\partial{\cal E}_{n,t}^{\cal U}}{\partial p_{n,m}} > 0 $, if $p_{n,m} > 0$. According to this finding, it is obvious that ${\cal E}_{n,t}^{\cal U}$ monotonically increases with $p_{n,m} > 0$. Consequently , we need to determine the range for $p_{n,m}$ according to the two constraints in problem $\mathbb{P}_4$ in order to find a minimum $p_{n,m}$ that minimizes the objective function Eq. (\ref{eq28}). The constraint $ p_n\in \left ( 0,p_n^{max} \right ]$ gives the first range limitation of $p_{n,m}$. According to the second constraint ${\cal T}_{n,t}\le \frac{\mathbb{I}}{\tau_e\tau_g}$ in problem $\mathbb{P}_4$, we can deform it by using ${\cal T}_{n,t} = {\cal T}_{n,t}^{\cal C} + {\cal T}_{n,t}^{\cal U}$ to derive another limitation:
\begin{equation}	
	{\cal T}_{n,t}^{\cal U} \le \frac{\mathbb{I}}{\tau_e\tau_g} - {\cal T}_{n,t}^{\cal C}.
\label{eq32}
\end{equation}
By plugging ${\cal T}_{n,t}^{\cal U} =\frac{\mathbb{Z}}{\mathbb{R}_{n,m}}$ into Eq. (\ref{eq32}), we have
\begin{equation}
	p_{n,m} \ge \frac{B_{n,m}^U\mathbb{N}_0\left ( 2^{\frac{\mathbb{Z}}{B_{n,m}^U\left ( \frac{\mathbb{I}}{\tau_e\tau_g}-{\cal T}_{n,t}^C  \right ) } }-1 \right ) }{h_{n,m}}.
\label{eq33}
\end{equation}
Let the RHS of Eq. (\ref{eq33}) be denoted as $ p_{n,\mathbb{I} }$, and combine the two restrictions of $p_{n,m}$, we can derive the optimal $p_{n,m}^*$, i.e.,
\begin{equation}
    p_{n,m}^*=\min\left \{{p_n^{max}} ,p_{n,\mathbb{I} }\right \}.
\end{equation}
Consequently, Theorem \ref{theorem2} is proved.

\subsection*{Supplement of Table 1}
\begin{figure*}[t]

\begin{center}
\subfigure[Initial Distribution]{
	\begin{minipage}[b]{.23\linewidth}
        \centering
        \includegraphics[scale=0.12]{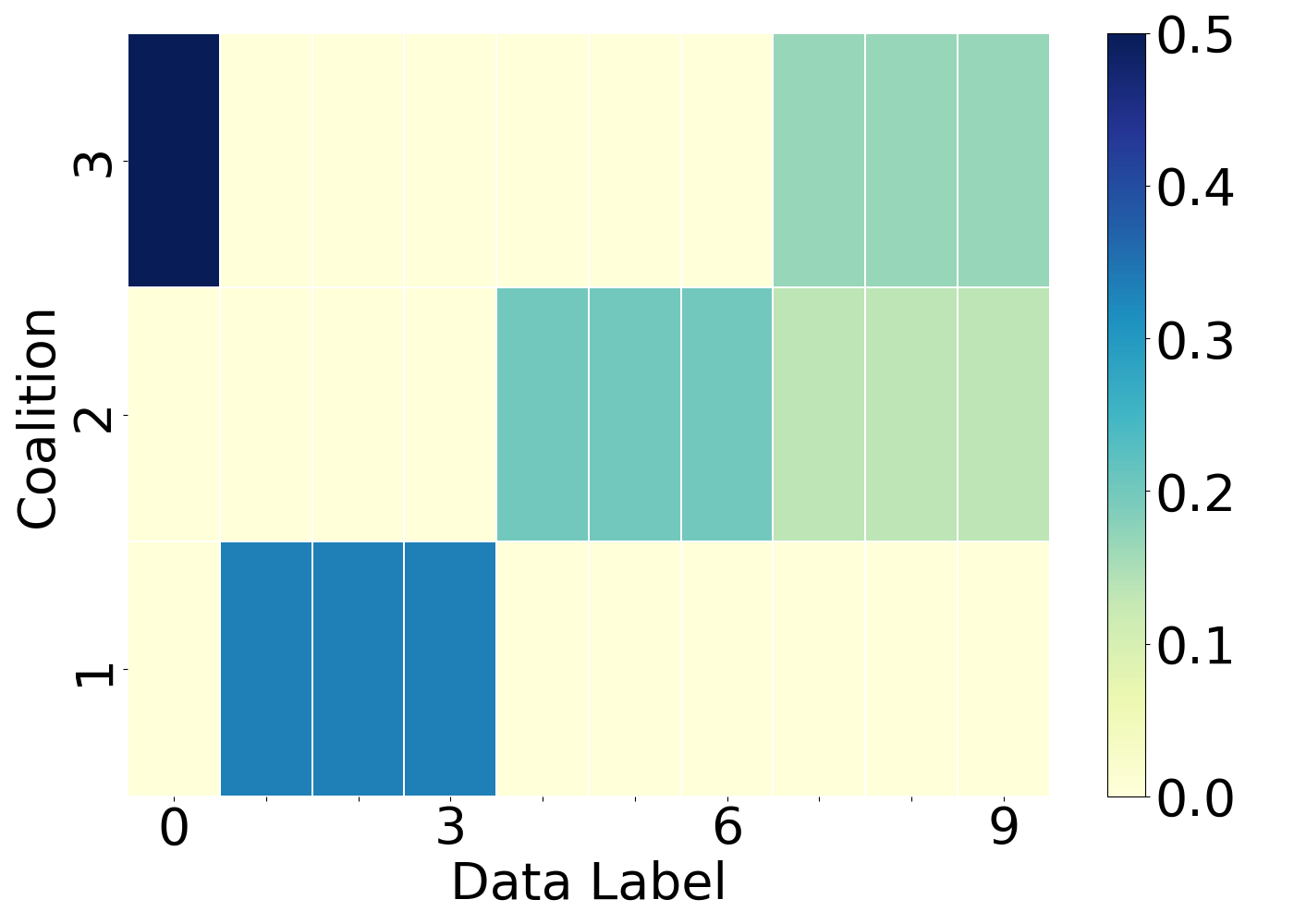}
 \vspace{-1em}
    \end{minipage}
\label{ma}
}
\subfigure[MA]{
	\begin{minipage}[b]{.23\linewidth}
        \centering
        \includegraphics[scale=0.12]{JSD_MA.png}
 \vspace{-1em}
    \end{minipage}
\label{ma1}
}
\subfigure[RH]{
	\begin{minipage}[b]{.23\linewidth}
        \centering
        \includegraphics[scale=0.12]{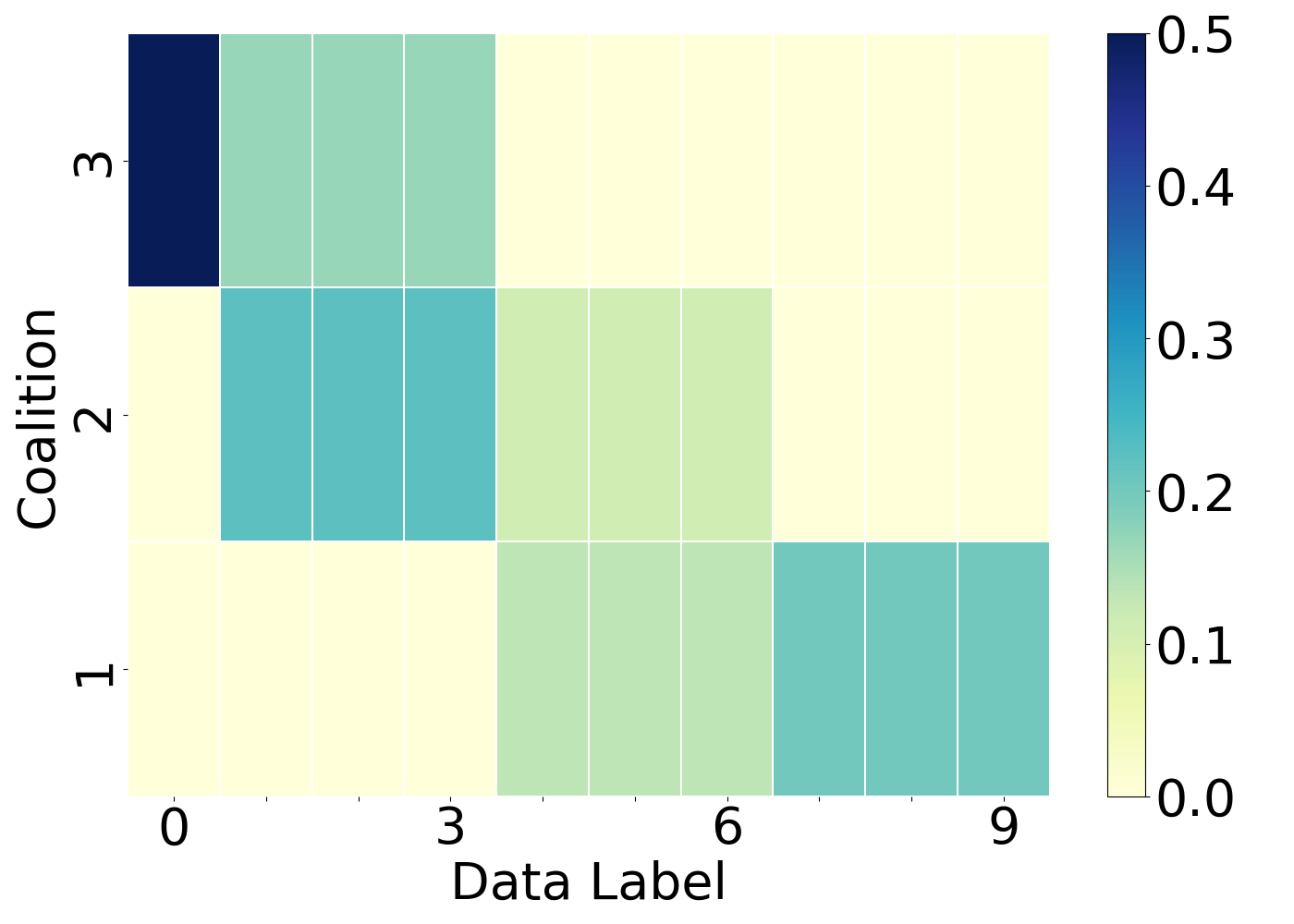}
 \vspace{-1em}
    \end{minipage}
\label{rh}
}
\subfigure[LEAP]{
	\begin{minipage}[b]{.23\linewidth}
        \centering
        \includegraphics[scale=0.12]{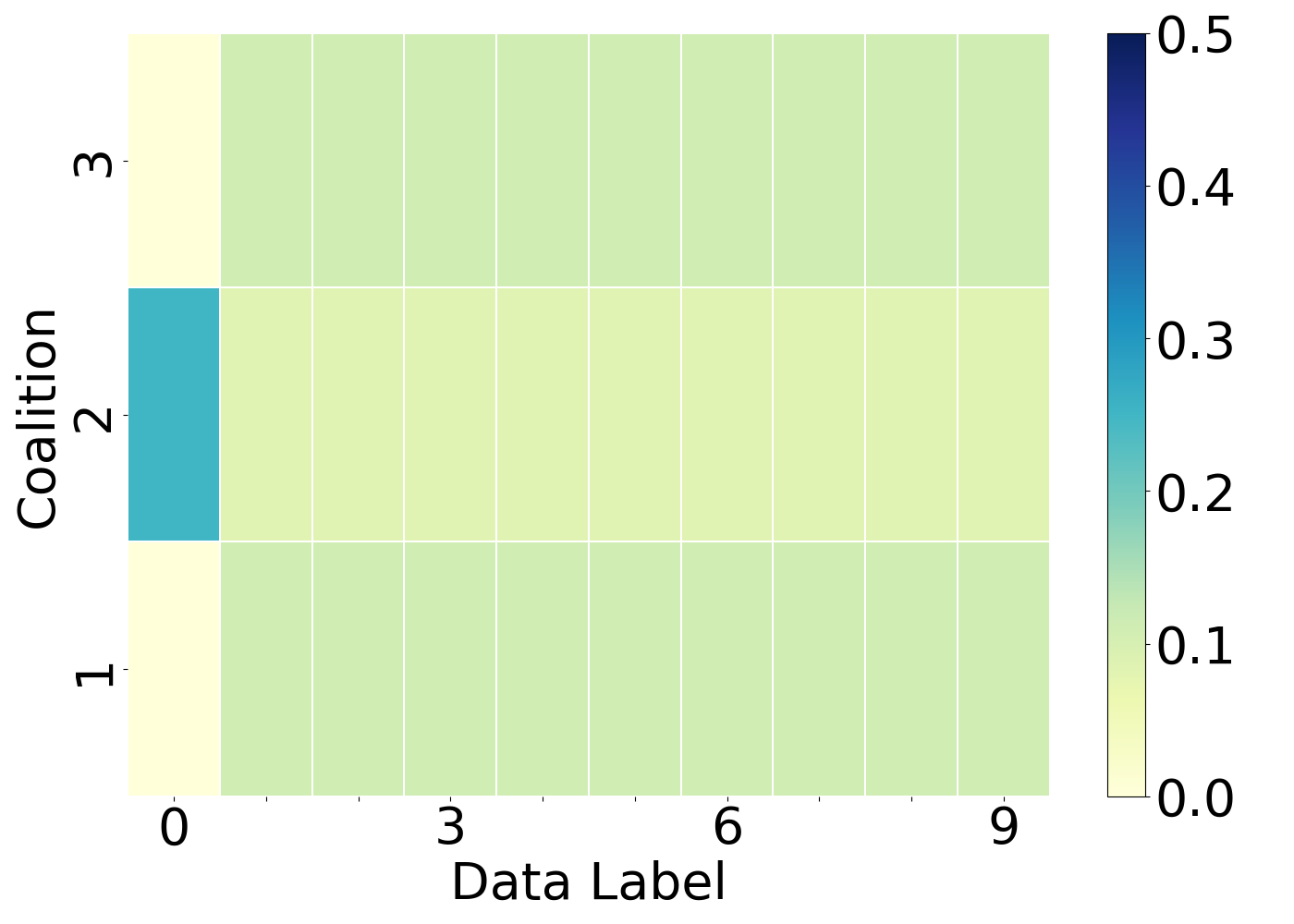}
 \vspace{-1em}
    \end{minipage}
\label{our}
}
 \vspace{-1em}
\caption{Variations in data distribution across three different methods.}
\label{555}
 \vspace{-0.6em}
\end{center}
 \vspace{-0.6em}
\end{figure*}

\begin{figure*}[t]
\begin{center}
\subfigure[MNIST]{
	\begin{minipage}[b]{.23\linewidth}
        \centering
        \includegraphics[scale=0.12]{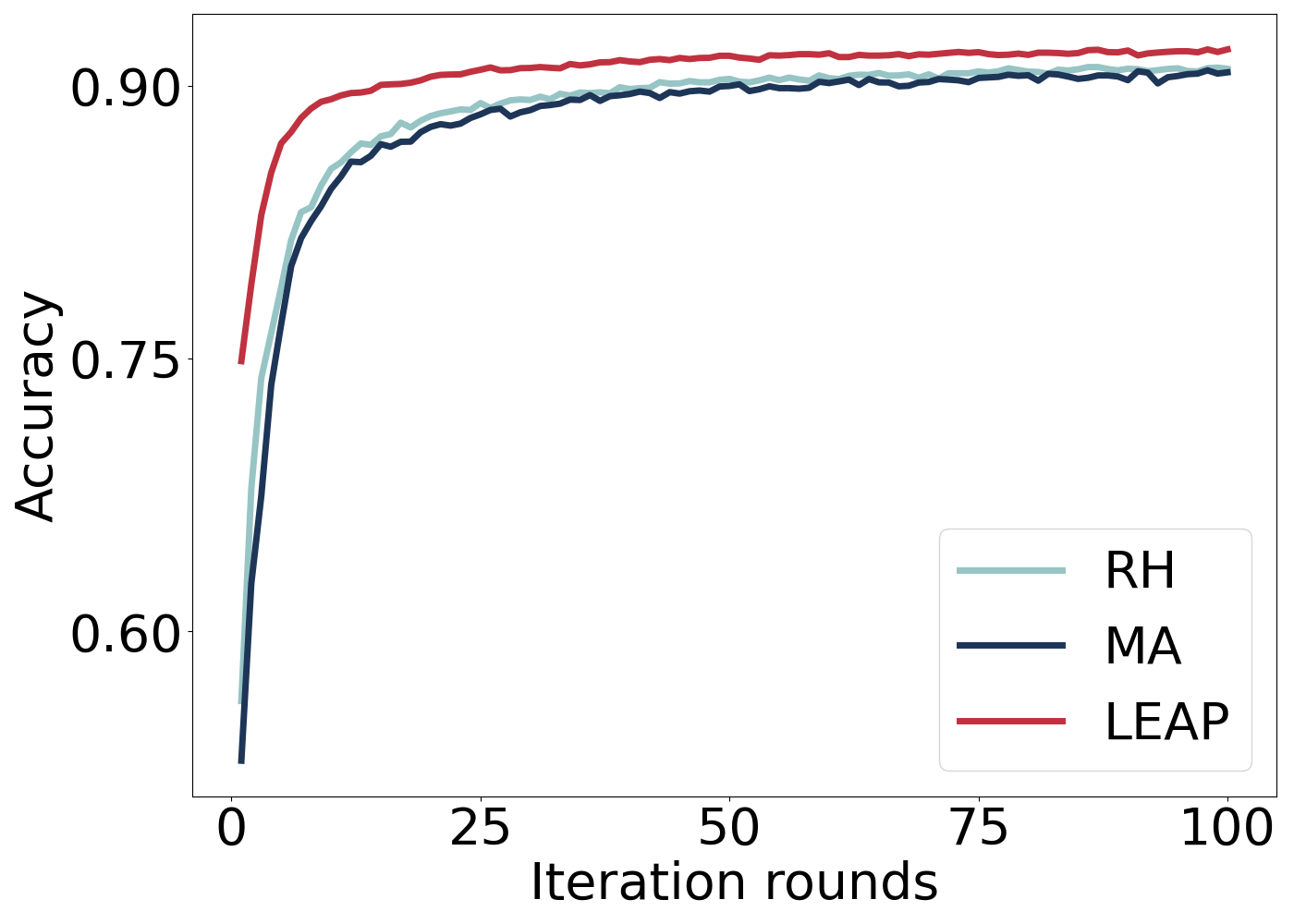}
 \vspace{-1em}
    \end{minipage}
}
\subfigure[CIFAR-10]{
	\begin{minipage}[b]{.23\linewidth}
        \centering
        \includegraphics[scale=0.12]{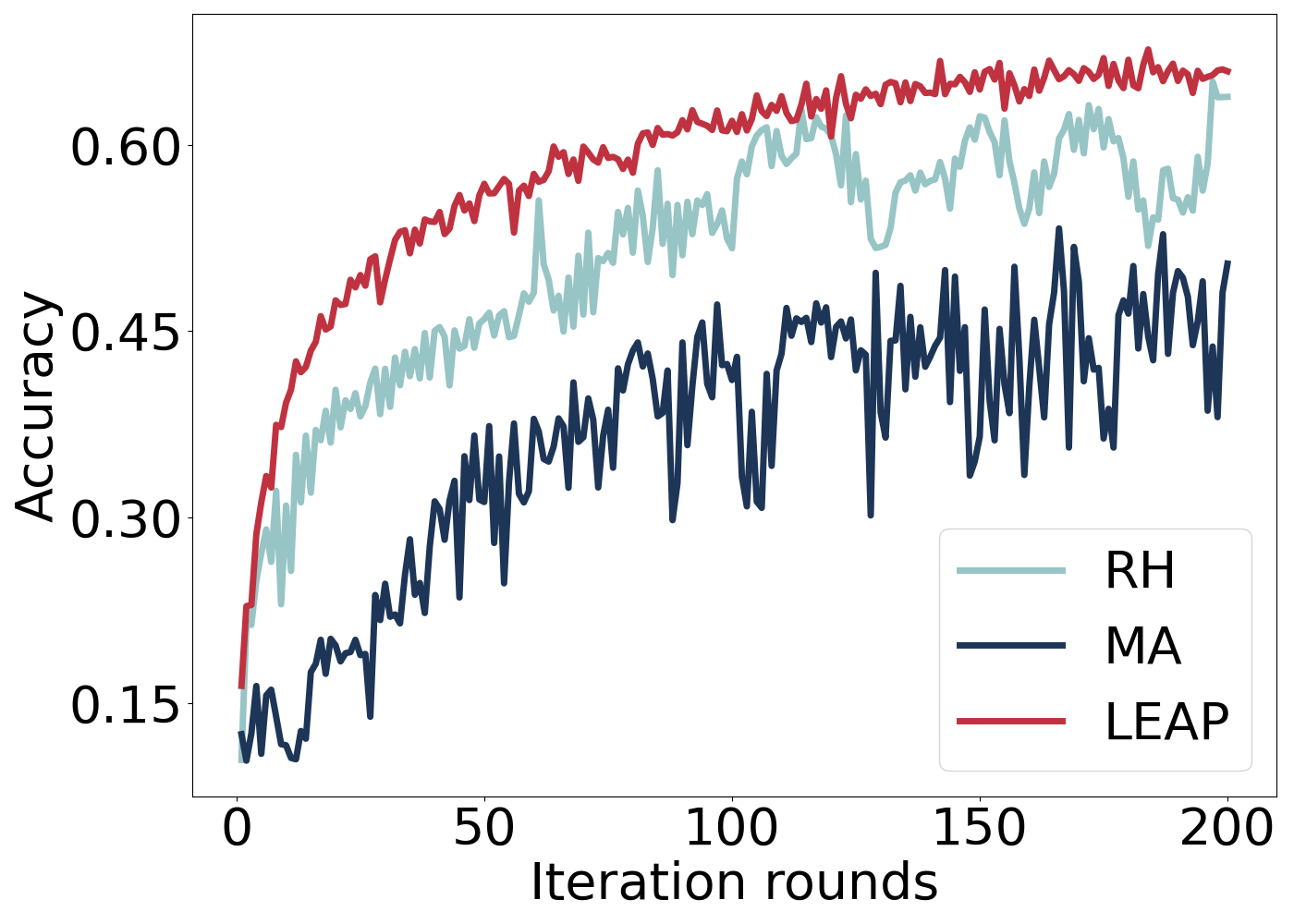}
 \vspace{-1em}
    \end{minipage}
}
\subfigure[SVHN]{
	\begin{minipage}[b]{.23\linewidth}
        \centering
        \includegraphics[scale=0.12]{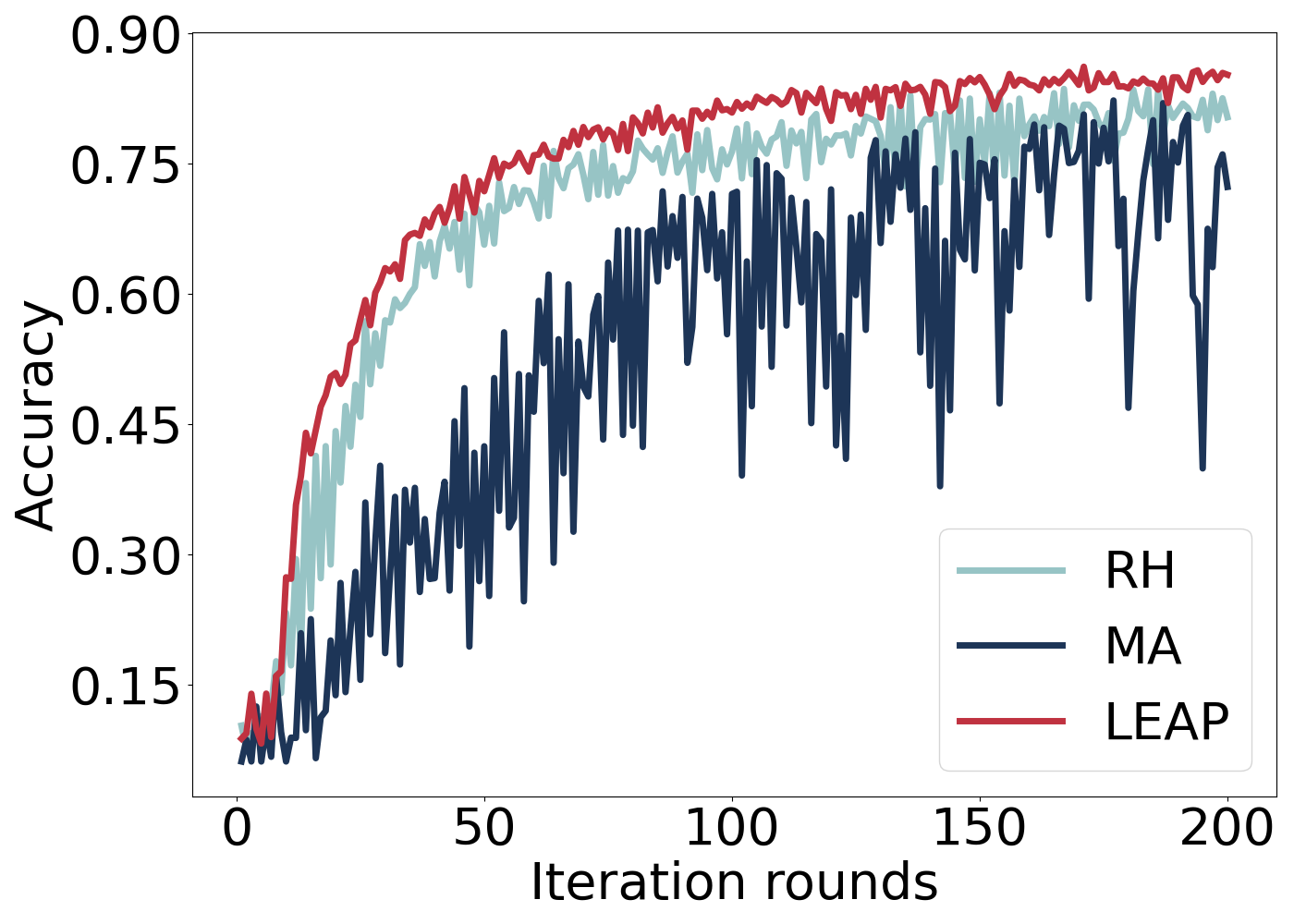}
 \vspace{-1em}
    \end{minipage}
}
\subfigure[CINIC-10]{
	\begin{minipage}[b]{.23\linewidth}
        \centering
        \includegraphics[scale=0.12]{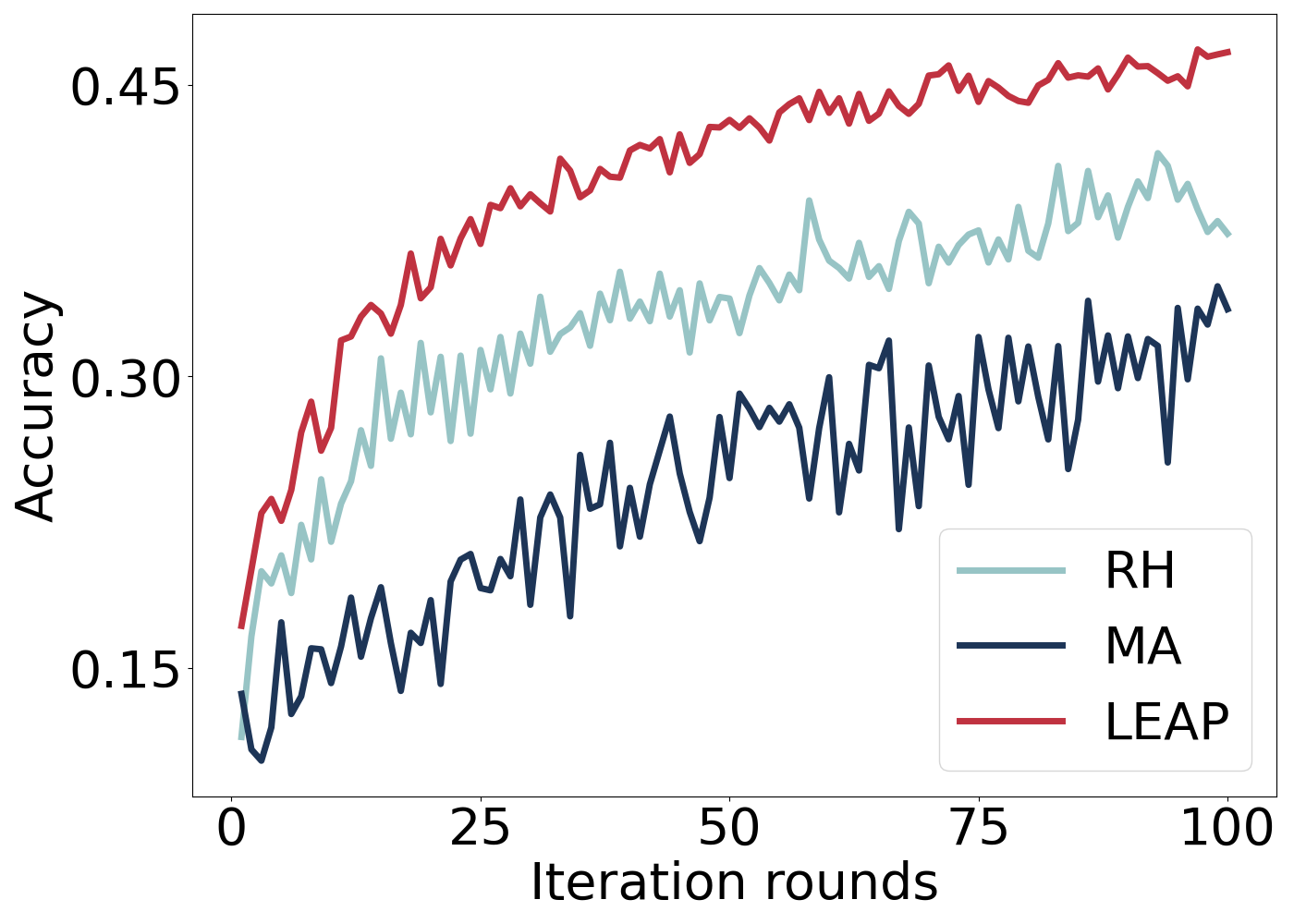}
 \vspace{-1em}
    \end{minipage}
}
 \vspace{-1em}
\caption{Global model performance comparison between LEAP and the state-of-the-art two baselines on four datasets.}
\label{666}
 \vspace{-0.6em}
\end{center}
 \vspace{-0.6em}
\end{figure*}

\begin{figure}[tbp]
\vspace{-0.6em}
\begin{center}
\subfigure[RH]{
	\begin{minipage}[t]{0.46\linewidth}
\centering
\includegraphics[scale=0.122]{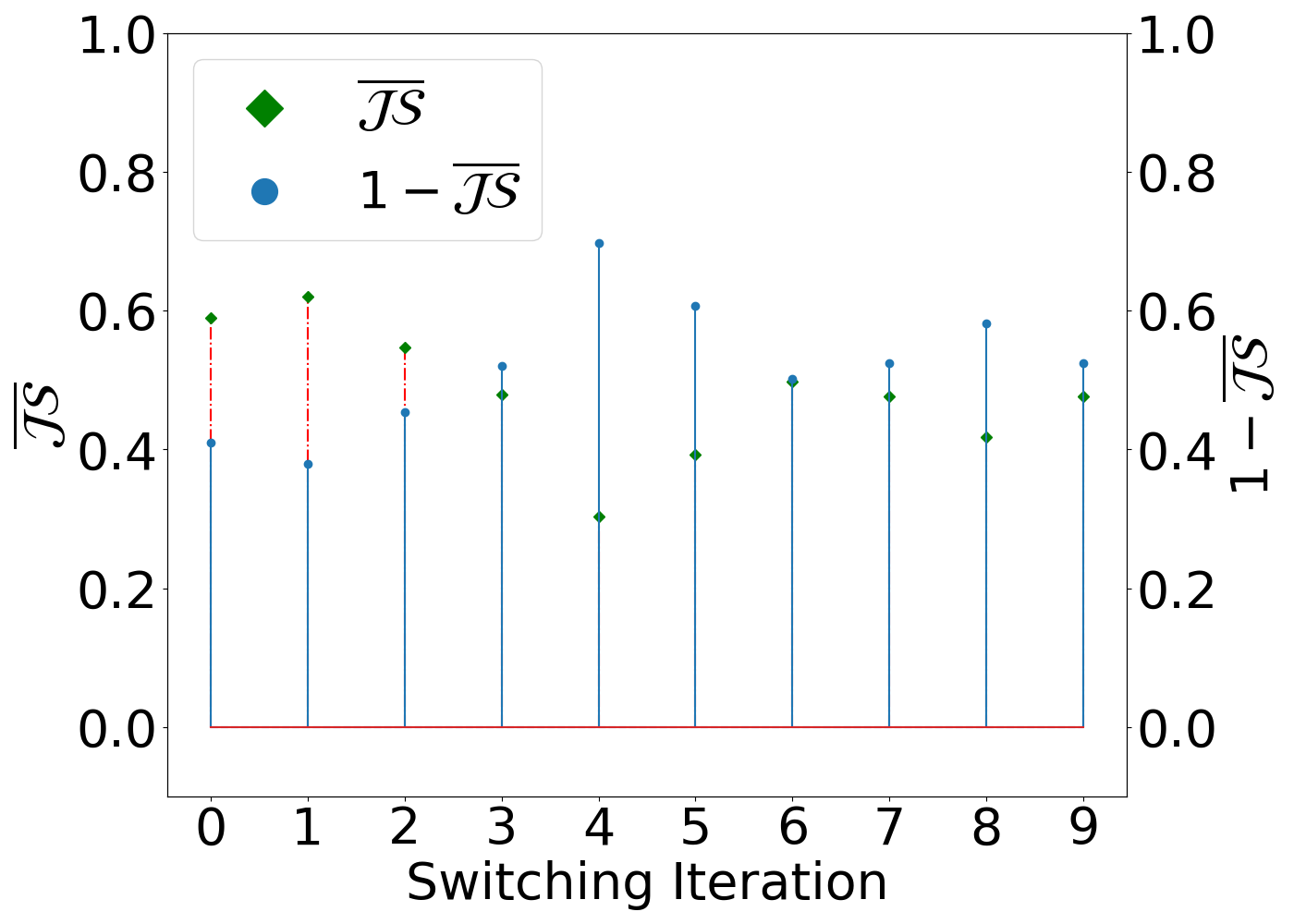}
\label{jsdrh}
\end{minipage}
}
\subfigure[LEAP]{
	\begin{minipage}[t]{0.46\linewidth}
\centering
\includegraphics[scale=0.122]{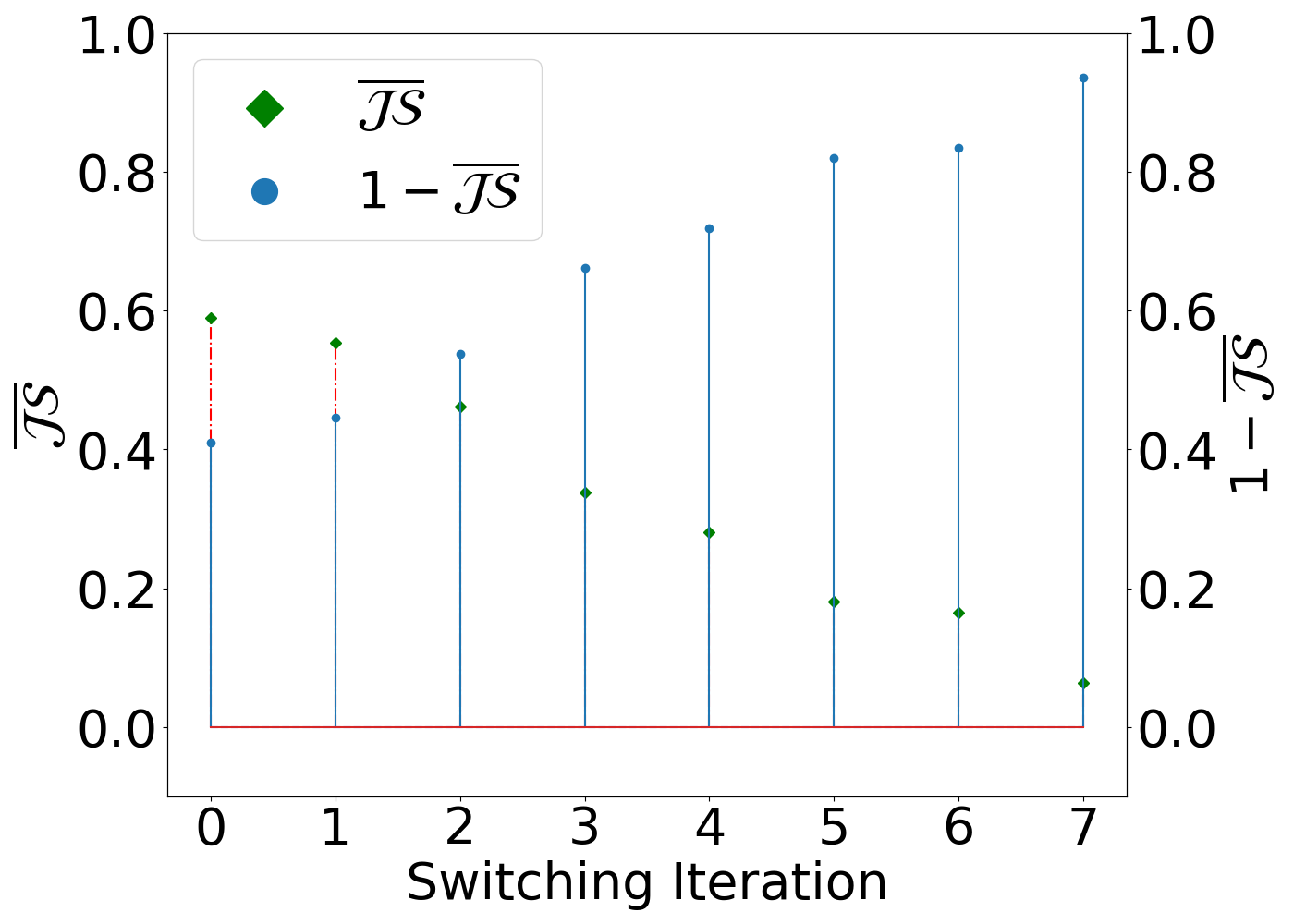}
\label{jsdg3}
\end{minipage}
}
\vspace{-1em}
\caption{Changes in the $\overline{\mathcal{JS}}$ value of RH and LEAP}
\label{777}
\end{center}
\vspace{-1em}
\end{figure}

In Table 1 of the submission file, we give the average model accuracy of RH, MA and LEAP on MNIST, CIFAR-10, SVHN and CINIC-10 datasets. For a better comparison, we additionally provide Figs. \ref{555} to \ref{777} as a supplement to Table \ref{table1}.

Firstly, Fig. \ref{555} shows the differences between the data distributions of different methods and the initial distribution. Fig. \ref{ma} shows the initial distribution of RH and our method before adjusting data distribution. Since the MA method does not involve the adjustment of data distribution, the state in Fig. \ref{ma1} is same as that in Fig. \ref{ma}. In Fig. \ref{rh}, the final data distribution of RH is different from the initial state. The main reason behind this phenomenon is that the RH does not perform coalition formation to optimize the data distribution. In contrast, as shown in Fig. \ref{our}, our results indicate that the data label distributions obtained by each coalition are ultimately the most similar.

Then, we give the results on four datasets based on the data distribution in Fig. \ref{555}. The results on the four datasets show that the distribution adjusted by our proposed method LEAP consistently yields optimal accuracy compared to RH and MA, with specific improvements of 10.75\% on the MNIST dataset, 22.23\% on the CIFAR-10 dataset, 20.7\% on the SVHN dataset, and 15.85\% on the CINIC-10 dataset, respectively. The main reason behind this phenomenon is that RA performs association formation with a selfish client preference rule without considering the impact on the coalition partition and data distribution, and MA discards some model parameters that fall below the loss threshold when performing aggregation based on marginal losses, resulting in data wasting. In contrast, LEAP performs coalition formation to achieve optimal data distribution among ESs to improve model performance without wasting data. It is worth noting that the average accuracy in Table \ref{table1} of the submission file is calculated by summing the data on the corresponding fold in Fig. \ref{666} and then dividing it by the number of global rounds.

Finally, we demonstrate the process of coalition formation using RH and LEAP separately to further compare them as shown in Fig. \ref{777}. There is no guarantee that $\overline{\mathcal{JS}}$ always reduces for each switch based on RH in Fig. \ref{jsdrh}. However, in Fig. \ref{jsdg3}, we can observe that each switch consistently achieves a lower $\overline{\mathcal{JS}}$ value when using LEAP. By combining Fig. \ref{rh} and Fig. \ref{our}, we can observe that LEAP has a better data distribution adjustment effect, which is consistent with the conclusion of our previous analysis.

\bibliographystyle{named}

\end{document}